\documentclass[12pt,reqno]{amsart}

\usepackage{amsmath,amsfonts,amsthm,cite,amssymb}

\usepackage{a4wide}
\usepackage[footnotesize,hang,sf]{caption}

\usepackage{titlesec}
\titleformat{\section}
{\bfseries\scshape\centering}
{\thesection.}{.5em}{}
\titleformat{\subsection}
{\rmfamily\bfseries}
{\thesubsection}{.5em}{}
\usepackage{times}
\newcommand{\vv}{\vec}
\usepackage{psfrag}
\usepackage[dvips]{graphicx}
\numberwithin{equation}{section}

\newtheorem{theorem}{Theorem}
\newtheorem{prop}[theorem]{Proposition}
\newtheorem{coro}[theorem]{Corollary}
\newtheorem{lemma}[theorem]{Lemma}
\newtheorem{conj}{Conjecture}
\DeclareMathOperator{\sign}{sign}
\DeclareMathOperator{\Tr}{Tr}

\DeclareMathOperator{\ch}{ch}

\DeclareMathOperator{\CS}{CS}

\DeclareMathOperator{\id}{id}

\begin{document}

\renewcommand{\thefootnote}{\fnsymbol{footnote}}
\baselineskip 15pt
\parskip 4pt
\sloppy




\title{
  Quantum Invariant, Modular Form, and Lattice Points
}


    \author{Kazuhiro \textsc{Hikami}}


  \address{Department of Physics, Graduate School of Science,
    University of Tokyo,
    Hongo 7--3--1, Bunkyo, Tokyo 113--0033,   Japan.
    }

    \urladdr{http://gogh.phys.s.u-tokyo.ac.jp/{\textasciitilde}hikami/}

    \email{\texttt{hikami@phys.s.u-tokyo.ac.jp}}


\date{August 24, 2004}

\begin{abstract}
We study the
Witten--Reshetikhin--Turaev SU(2) invariant for the
Seifert manifold with 4-singular fibers.
We define the Eichler integrals of the modular forms with half-integral
weight, and we show that the  invariant is rewritten as a sum of the
Eichler integrals.
Using a nearly modular property of the Eichler integral, 
we give an exact asymptotic expansion of the WRT invariant
in $N\to\infty$.
We  reveal that the number of dominating
terms, which is the number  of the non-vanishing
Eichler integrals in a limit $\tau\to N\in\mathbb{Z}$,
is
related to that  of lattice points inside 4-dimensional simplex,
and we discuss a relationship with the irreducible representations
of the fundamental group.

\end{abstract}





\maketitle

\section{Introduction}

The    SU(2) Witten invariant~\cite{EWitt89a} for 3-manifold
$\mathcal{M}$ is defined by
\begin{equation}
  Z_k(\mathcal{M})
  =
  \int \exp \Bigl(
    2 \, \pi \, \mathrm{i} \, k \,  \CS(A)
  \Bigr) \,
  \mathcal{D} A
\end{equation}
where
$k\in\mathbb{Z}$, and $\CS(A)$ is the Chern--Simons integral
\begin{equation}
  \CS(A)
  =
  \frac{1}{8 \, \pi^2}
  \int_{\mathcal{M}}
  \Tr
  \left(
    A \wedge \mathrm{d} A
    + \frac{2}{3}
    \, A \wedge A \wedge A
  \right)
\end{equation}
This invariant can be constructed rigorously using the quantum
invariants of framed link as the Reshetikhin--Turaev invariant
$\tau_N(\mathcal{M})$~\cite{ResheTurae91a},
which
is normalized
as
\begin{equation}
  Z_k(\mathcal{M})
  =
  \frac{\tau_{k+2}(\mathcal{M})}{
    \tau_{k+2} \left( S^2 \times S^1 \right)}
\end{equation}
where
\begin{gather*}
  \tau_N \left( S^2 \times S^1 \right)
  =
  \sqrt{\frac{N}{2}} \,
  \frac{1}{
    \sin(\pi/N)
  }
\end{gather*}
and we have
\begin{equation*}
  \tau_N  \left( S^3  \right)=1  
\end{equation*}

Studies on these quantum invariants have been extensively developed.
Recently pointed out was a close relationship between the
Witten--Reshetikhin--Turaev (WRT)
invariant and modular form with half-integral weight.
In Ref.~\citen{LawrZagi99a},
the WRT invariant for the Poincar{\'e} homology sphere $\Sigma(2,3,5)$
was identified with the Eichler integral of the modular form with weight $3/2$.
This result was further developed in Ref.~\citen{KHikami04b} where
properties of the
WRT invariant for the Brieskorn homology sphere $\Sigma(p,q,r)$
were investigated
(see also Refs.~\citen{KHikami02c,KHikami03a,KHikami03c,KHikami02b},
in which clarified were the similar structures
of the colored Jones polynomial for torus knots and links).
One of benefits of the correspondence between the quantum invariant
and the modular form is that
we can obtain
the exact  asymptotic expansion  from the modular property
(see
Refs.~\citen{FreeGomp91a,LCJeff92a,RLawre95a,Rozan95a,Rozan96c,Rozan96e,LRozan97a,LawreRozan99a}
for studies of  asymptotic behavior of the SU(2) WRT invariant
by different manner).
We can also find that the number of the non-vanishing Eichler integrals in 
a limit $\tau \to N\in\mathbb{Z}$ coincides with
that of  the integral lattice points
inside the 3-dimensional tetrahedron.
We can then reinterpret
the topological invariants
such as the Casson invariant, the Reidemeister torsion,
and  the
Chern--Simons invariant
from the viewpoint of the modular form.

\begin{figure}[htbp]
  \centering
  
    \begin{psfrags}
    \psfrag{x}{$0$}
    \psfrag{a}{$p_1/q_1$}
    \psfrag{b}{$p_2/q_2$}
    \psfrag{c}{$p_3/q_3$}
    \psfrag{d}{$p_4/q_4$}
    \includegraphics[scale=1.4, bb=-60 -50 60 50]{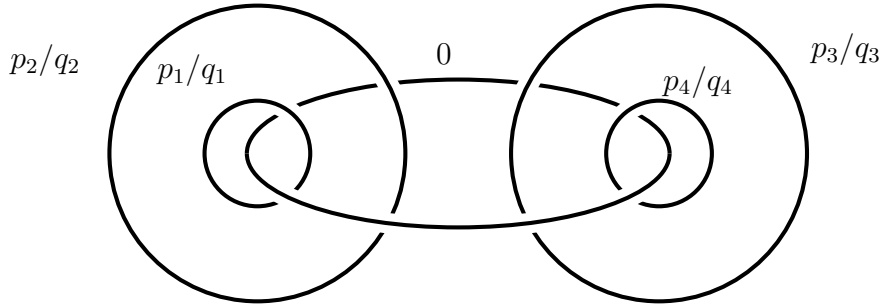}
  \end{psfrags}
  \caption{
    Rational surgery description of the Seifert manifold
    $\Sigma(p_1, p_2, p_3, p_4)$ with 4-singular fibers}
  \label{fig:4-fiber}
\end{figure}

Purpose of this paper is to continue studies in
Ref.~\citen{KHikami04b}, and to reveal a close connection between the
WRT invariant for the Seifert manifold
and the Eichler integral of the modular form.
We especially study the Seifert manifold with 4-singular fibers
$\Sigma(\vv{p}) = \Sigma(p_1,p_2,p_3, p_4)$
where $p_j$ are pairwise coprime integers.
This manifold has a rational surgery description as in
Fig.~\ref{fig:4-fiber},
and the fundamental group has the presentation
\begin{equation}
  \label{fundamental_group}
  \pi_1  \bigl(
    \Sigma(p_1,p_2,p_3,p_4)
  \bigr)
  =
  \left\langle
    x_1, x_2, x_3, x_4, h
    ~\bigg|~
    \begin{array}{c}
      \text{$h$  center}
      \\[1mm]
      \text{$x_k^{~p_k} = h^{-q_k}$ for $k=1,2,3,4$}
      \\[1mm]
      x_1 \, x_2 \, x_3 \, x_4 =1
    \end{array}
  \right\rangle
\end{equation}
where $q_k \in \mathbb{Z}$ such that
\begin{equation}
  P \sum_{j=1}^4 \frac{q_j}{p_j} = 1
\end{equation}
Here and hereafter we use
\begin{equation*}
  P \equiv P(p_1, p_2, p_3, p_4)
  = \prod_{j=1}^4 p_j
\end{equation*}

This article is organized as follows.
In Section~\ref{sec:preparation} we briefly discuss construction of the WRT
invariant for the Seifert manifold.
We then prepare    vector modular forms
with half-integral weight,
and we consider a nearly modular property of the Eichler integral.
In Section~\ref{sec:invariant_Eichler} we prove that the WRT invariant
for the Seifert manifold is written as a sum of the Eichler integrals of two types of the half-integral weight modular forms.
Using a nearly modular property of the Eichler integrals, we  obtain
the exact asymptotic expansion of the WRT invariant in
Section~\ref{sec:asymptotic}.
We pay attention to dominating exponential factors of the WRT
invariant, and  show that
the number of the non-vanishing terms is related to that of the irreducible representations of the fundamental group $\pi_1(\mathcal{M})$,
and that
they give the Chern--Simons invariant.
We also compute  the Ohtsuki invariant number-theoretically by use  of
the exact asymptotic expansion formula.
In Section~\ref{sec:example} we take some examples in detail,
and compare numerically our asymptotic formula with the exact value of the
quantum invariant.
The last section is devoted to discussions.

\section{Preliminaries}
\label{sec:preparation}
\subsection{The Witten--Reshetikhin--Turaev Invariant of the Seifert
  Manifold}
\label{sec:WRT}

We compute the SU(2) WRT invariant for the
Seifert homology manifold $\Sigma(p_1, p_2, p_3, p_4)$
with 4-singular fibers.
In general the WRT invariant
$\tau_N(\mathcal{M})$
for  3-manifold $\mathcal{M}$
can be constructed  based
on a surgery 
description on framed link.
In our case, we need the colored Jones polynomial for a link depicted
in Fig.~\ref{fig:4-fiber}.
To construct  the colored Jones polynomial
for this link, we recall a fact that,
when a link $\mathcal{L}$  is composed from three knots
$\mathcal{K}_{0,1,2}$,
the colored Jones polynomial for $\mathcal{L}$ is given by
\begin{equation*}
  J_{k_0, k_1, k_2} (\mathcal{K}_0 \cup \mathcal{K}_1 \cup \mathcal{K}_2)
  =
  \frac{J_{k_0, k_1} (\mathcal{K}_0 \cup \mathcal{K}_1) \,
    J_{k_0, k_2}(\mathcal{K}_0 \cup \mathcal{K}_2)
  }{
    J_{k_0}(\mathcal{K}_0)
  }
\end{equation*}
where $k_a$ denotes a color of knot $\mathcal{K}_a$.
Using this property and  an explicit form of the colored Jones invariant for the Hopf link,
we see that the colored Jones polynomial for a link $\mathcal{L}$
in
Fig.~\ref{fig:4-fiber} is given by
\begin{equation}
  J_{k_0,k_1,k_2,k_3,k_4}(\mathcal{L})
  =
  \frac{1}{\sin ( \pi /N )} \cdot
  \frac{
    \prod_{j=1}^4 \sin
    \left(
      k_0 \, k_j \, \pi/N
    \right)
  }{
    \bigl(
    \sin
    ( k_0 \, \pi /N )
    \bigr)^3
  }
\end{equation}
Here $k_{j>0}$ is a color for knot which is to be $p_j/q_j$-surgery,
and $k_0$ denotes a color for knot having a linking number 1 with
other knots.
We then apply a rational surgery formula presented in
Ref.~\citen{LCJeff92a},
and we finally obtain the following result.
See Ref.~\citen{LawreRozan99a}
(also Ref.~\citen{KHikami04b})
for detail of computations.

\begin{prop}[\cite{LawreRozan99a}]
  The WRT invariant
  $\tau_N(\mathcal{M})$
  for the Seifert manifold
  $\mathcal{M}=\Sigma(p_1, p_2, p_3,  p_4)$
  is given by
  \begin{multline}
    \label{Rozansky}
    \mathrm{e}^{
      \frac{2 \pi \mathrm{i}}{N}
      ( \frac{\phi}{4} - \frac{1}{2} )
    } \,
    \left(
      \mathrm{e}^{\frac{2 \pi \mathrm{i}}{N}} - 1
    \right)
    \,    \tau_N \bigl( \Sigma(p_1,p_2,p_3, p_4) \bigr)
    \\
    =
    \frac{\mathrm{e}^{\pi \mathrm{i}/4}}{2 \, \sqrt{2 \, P \, N}}
    \,
    \sum_{\substack{
        n= 0 \\
        N \, \nmid \, n
    }}^{2 \, P \, N-1 }
    \mathrm{e}^{- \frac{1}{2 P N} n^2 \pi \mathrm{i} } \,
    \frac{
      \prod_{j=1}^4
      \left(
        \mathrm{e}^{\frac{  n}{N p_j} \pi \mathrm{i}}
        -
        \mathrm{e}^{- \frac{ n}{N p_j} \pi \mathrm{i} }
      \right)
    }{
      \left(
        \mathrm{e}^{\frac{n}{N} \pi \mathrm{i}}
        -
        \mathrm{e}^{ - \frac{ n}{N} \pi \mathrm{i}}
      \right)^2
    }
  \end{multline}
  where
  \begin{equation}
    \label{phi_definition}
    \phi
    \equiv
    \phi(p_1, p_2, p_3,p_4)
    =
    3 - \frac{1}{P}
    +
    12 \,
    \sum_{j=1}^4
    s
    \left( \frac{P}{p_j}, p_j  \right)
  \end{equation}
\end{prop}

Here we have used  the Dedekind sum
(see, \emph{e.g.},  Ref.~\citen{RademGross72})
defined by
\begin{equation}
  \label{Dedekind_sum}
  s(b,a) =
  \sum_{k=1}^{a}
  \Bigl(\Bigl( \frac{  \ k \  }{a} \Bigr) \Bigr) \cdot
  \Bigl(\Bigl( \frac{  \  k \, b   \  }{a} \Bigr) \Bigr) 
\end{equation}
with  coprime integers $a \geq 1$ and $b$, and
$((x))$ is the sawtooth function
\begin{equation*}
  ((x))
  =
  \begin{cases}
    x - \lfloor x \rfloor - \frac{1}{2}
    &
    \text{if $x \not\in \mathbb{Z}$}
    \\[3mm]
    0
    &
    \text{if $x \in \mathbb{Z}$}
  \end{cases}
\end{equation*}
where $\lfloor x \rfloor$ is the greatest integer not exceeding $x$.
It is known that the Dedekind sum is rewritten as the cotangent sum
\begin{equation*}
  s(b,a)
  =
  \frac{1}{4 \, a}
  \sum_{k=1}^{a-1}
  \cot \left( \frac{  \ k \   }{a}  \, \pi \right) \,
  \cot \left( \frac{  \  k \, b  \  }{a} \, \pi \right) 
\end{equation*}
and that  the modular property of
the logarithm of the Dedekind $\eta$-function
$ \eta(\tau)$
gives
the reciprocity formula
\begin{equation*}
  s(b, a) + s(a, b)
  =
  -\frac{1}{4} + \frac{1}{12} \,
  \left(
    \frac{a}{b} + \frac{b}{a} +
    \frac{1}{a \, b}
  \right)
\end{equation*}
It is noted that
\begin{gather*}
  s(-b, a)= - s (b,a)
  \\[2mm]
  s(b,a) = s(c ,a )
  \qquad
  \text{for $b \, c = 1 \pmod a$}
\end{gather*}

\subsection{Modular Form}
\label{sec:modular}

We fix $\vv{p}=(p_1, p_2, p_3, p_4)$ where
$p_j$ are pairwise coprime positive integers.
For a quadruple
$\vv{\ell}=(\ell_1, \ell_2, \ell_3, \ell_4) \in \mathbb{Z}^4$
satisfying
$0<\ell_j<p_j$,
we define even periodic functions
$\chi_{2 P}^{\vv{\ell}}(n)
=\chi_{2 P}^{(\ell_1, \ell_2, \ell_3, \ell_4)}(n)$
with modulus $2 \, P$ as
\begin{equation}
  \chi_{2 P}^{\vv{\ell}}(n)
  =
  \begin{cases}
    \displaystyle
    - \prod_{j=1}^4 \varepsilon_j
    &
    \text{if
      $\displaystyle
      n \equiv
      P \, \left(
        1 + \sum_{j=1}^4 \varepsilon_j \,
        \frac{\ell_j}{p_j}
      \right)
      \mod 2 \, P
      $
    }
    \\[2mm]
    0 &
    \text{others}
  \end{cases}
\end{equation}
where $\varepsilon_j \in \{ 1, -1 \} $ for $\forall j$.
There are $2^4=16$ non-zero
$\chi_{2 P}^{\vv{\ell}}(n)$
taking values $\pm 1$
for $0<n<2 \, P$, and we have a mean value zero,
\begin{equation}
  \sum_{n=0}^{2 P -1}
  \chi_{2 P}^{\vv{\ell}}(n) =0
\end{equation}

We define an involution
\begin{gather}
  \sigma_i(\vv{\ell})
  =
  (\ell_1, \dots, p_i - \ell_i, \dots, p_4)
  \\[2mm]
  \begin{aligned}
    \sigma_{i j}(\vv{\ell})
    & \equiv
    \sigma_i \circ \sigma_j (\vv{\ell})
    \\
    &    =
    (\ell_1,\dots, p_i - \ell_i, \dots, p_j - \ell_j, \dots, p_4)
  \end{aligned}
\end{gather}
for $1 \leq i \neq j \leq 4$.
We see that the periodic function
$\chi_{2  P}^{\vv{\ell}}(n)$
is invariant under actions of $\sigma_{i j}$ and
$\sigma_{1 2}\circ\sigma_{3 4}$;
\begin{equation}
  \label{invariant_chi}
  \begin{aligned}
    \chi_{2 P}^{\vv{\ell}}(n)
    & =
    \chi_{2 P}^{\sigma_{i j}(\vv{\ell})}(n)
    \\
    & =
    \chi_{2 P}^{\sigma_{1 2} \circ \sigma_{3 4} (\vv{\ell})}(n)
  \end{aligned}
\end{equation}
We note that
\begin{equation}
  \label{chi_sigma}
  \chi_{2 P}^{\vv{\ell}}(n+P)
  = -
  \chi_{2 P}^{\sigma_i(\vv{\ell})}(n)
\end{equation}

By means of the periodic functions $\chi_{2 P}^{\vv{\ell}}(n)$,
we define the $q$-series by
\begin{equation}
  \Phi_{\vv{p}}^{\vv{\ell}}(\tau)
  =
  \frac{1}{2}
  \sum_{n \in \mathbb{Z}}
  \chi_{2 P}^{\vv{\ell}}(n)
  \,
  q^{\frac{n^2}{4  \, P}}
\end{equation}
where as usual we have
\begin{equation*}
  q=\exp \left(2 \, \pi \, \mathrm{i} \, \tau \right)
\end{equation*}
for $\tau$   in the upper half plane, $\tau \in \mathbb{H}$.
Due to the symmetry of $\chi_{2 P}^{\vv{\ell}}(n)$
under involutions~\eqref{invariant_chi},
the number of the independent functions
$\Phi_{\vv{p}}^{\vv{\ell}}(\tau)$
is given by
\begin{equation}
  D\equiv
  D(p_1, p_2, p_3 , p_4 )
  =
  \frac{1}{8} \,
  \prod_{j=1}^4
  (p_j - 1 )
  \label{define_D}
\end{equation}

This set of functions $\Phi_{\vv{p}}^{\vv{\ell}}(\tau)$
is a $D$-dimensional vector   modular
form 
with weight $1/2$;
applying the Poisson summation formula
\begin{equation}
  \label{Poisson_summation}
  \sum_{n \in \mathbb{Z}} f(n)
  =
  \sum_{n \in \mathbb{Z}}
  \int_{-\infty}^\infty
  \mathrm{e}^{-2 \,  \pi \,  \mathrm{i} \,  n \, t} \,
  f(t) \,
  \mathrm{d} t
\end{equation}
we find under the $S$- and $T$-modular transformations that
\begin{gather}
  \label{Phi_under_S}
  \Phi_{\vv{p}}^{\vv{\ell}}(\tau)
  =
  \sqrt{\frac{  \  \mathrm{i}  \  }{\tau}} \,
  \sum_{
    \ell_1^\prime,
    \ell_2^\prime,
    \ell_3^\prime,
    \ell_4^\prime,
  }
  \mathbf{S}_{\vv{\ell}}^{\vv{\ell^\prime}}
  \,
  \Phi_{\vv{p}}^{\vv{\ell^\prime}}
  (-1/\tau)
  \\[2mm]
  \Phi_{\vv{p}}^{\vv{\ell}}(\tau+1)
  =
  \mathbf{T}^{\vv{\ell}}
  \,
  \Phi_{\vv{p}}^{\vv{\ell}}(\tau)
\end{gather}
where a sum of quadruples
$\vv{\ell^\prime}=(\ell_1^\prime, \ell_2^\prime, \ell_3^\prime,
\ell_4^\prime)$ runs over $D$-dimensional space, and explicit forms of the
$\mathbf{S}$ and $\mathbf{T}$ matrices are  respectively
given by
\begin{gather}
  \label{S_matrix}
  \mathbf{S}_{\vv{\ell}}^{\vv{\ell^\prime}}
  =
  \frac{16}{\sqrt{2 \, P}} \,
  (-1)^{
    P
    \left(
      1 + \sum_{j=1}^4 \frac{\ell_j+\ell_j^\prime}{p_j}
    \right)
    +
    P \sum_j \sum_{k \neq j }
    \frac{\ell_j \, \ell_k^\prime}{
      p_j  \, p_k
    }
  } \,
  \prod_{j=1}^4
  \sin
  \left(
    P \, \frac{\ell_j \, \ell_j^\prime}{p_j^{~2}} \,\pi
  \right)
  \\[2mm]
  \mathbf{T}^{\vv{\ell}}
  =
  \exp
  \left(
    \frac{P}{2} \,
    \Bigl(
      1+ \sum_{j=1}^4 \frac{\ell_j}{p_j}
    \Bigr)^2 \,
    \pi \, \mathrm{i}
  \right)
  \label{T_matrix}
\end{gather}

We further introduce other modular functions.
We set
\begin{equation}
  \Psi_P^{(a)}(\tau)
  =
  \frac{1}{2}
  \sum_{n \in \mathbb{Z}}
  n \, \psi_{2 P}^{(a)}  (n) \,
  q^{\frac{n^2}{4 \, P}}
\end{equation}
where $a\in\mathbb{Z}$ and
$ 0 < a < P$, and
$\psi_{2 P}^{(a)}(n)$ is the odd periodic function
\begin{equation}
  \psi_{2 P}^{(a)}(n)
  =
  \begin{cases}
    \pm 1 &
    \text{for $n \equiv \pm a \mod 2 \, P$}
    \\[2mm]
    0 &
    \text{others}
  \end{cases}
\end{equation}
By use of the Poisson summation formula~\eqref{Poisson_summation}
we see 
that the function $\Psi_P^{(a)}(\tau)$ is 
the ($P-1$)-dimensional vector modular form with
weight $3/2$
satisfying
\begin{gather}
  \label{Psi_under_S}
  \Psi_P^{(a)}(\tau)
  =
  \left(
    \frac{  \  \mathrm{i}  \  }{\tau}
  \right)^{3/2} \,
  \sum_{b=1}^{P-1}
  \mathbf{M}_b^a \,
  \Psi_P^{(b)}(-1/\tau)
  \\[2mm]
  \Psi_P^{(a)}(\tau+1)
  =
  \exp
  \left(
    \frac{a^2}{2 \, P} \, \pi \, \mathrm{i}
  \right) \,
  \Psi_P^{(a)}(\tau)
\end{gather}
where $\mathbf{M}$ is a $(P-1)\times (P-1)$ matrix defined by
\begin{equation}
  \mathbf{M}_b^a =
  \sqrt{\frac{  \  2  \  }{P}} \,
  \sin
  \left(\frac{a \, b}{P} \, \pi \right)
\end{equation}
It should be remarked that
we have
\begin{equation*}
  \Psi_{P=2}^{(1)}(\tau)
  = \bigl(
  \eta(\tau)
  \bigr)^3
\end{equation*}
and that
the character of the affine Lie algebra $\widehat{su}(2)_{P-2}$
is given by
(see \emph{e.g.} Ref.~\citen{Kac90})
\begin{equation*}
  \ch_a(\tau) =
  \frac{\Psi_P^{(a)}(\tau)}{
    \bigl(
      \eta(\tau)
    \bigr)^3
  }
\end{equation*}

\subsection{Eichler Integral}
\label{sec:Eichler}

Following Refs.~\citen{LawrZagi99a,DZagie01a}
we define the Eichler integrals of the
half-integral weight modular forms
$\Phi_{\vv{p}}^{\vv{\ell}}(\tau)$ and $\Psi_P^{(a)}(\tau)$
as follows;
\begin{gather}
  \widetilde{\Phi}_{\vv{p}}^{\vv{\ell}}(\tau)
  =
  \sum_{n=0}^\infty
  n \, \chi_{2 P}^{\vv{\ell}}(n) \,
  q^{\frac{n^2}{4  \, P}}
  \\[2mm]
  \widetilde{\Psi}_P^{(a)}(\tau)
  =
  \sum_{n=0}^\infty
  \psi_{2 P}^{(a)}(n) \,
  q^{\frac{n^2}{4  \, P}}
\end{gather}
which are defined for $\tau \in \mathbb{H}$.
Limiting values of these Eichler integrals
are given in the following Proposition.

\begin{prop}
  Limiting values of the Eichler integrals
  $\widetilde{\Phi}_{\vv{p}}^{\vv{\ell}}(\tau)$
  and
  $\widetilde{\Psi}_P^{(a)}(\tau)$
  at $\tau \to
  \frac{M}{N} \in \mathbb{Q}$
  are  respectively given by
  \begin{gather}
    \widetilde{\Phi}_{\vv{p}}^{\vv{\ell}}(M/N)
    =
    -P \, N
    \sum_{k=1}^{2 P N}
    \chi_{2 P}^{\vv{\ell}}(k) \,
    \mathrm{e}^{\pi \mathrm{i} \frac{M}{N} \frac{k^2}{2 P}} \,
    B_2
    \left(
      \frac{k}{2 \, P \, N}
    \right)
    \\[2mm]
    \widetilde{\Psi}_P^{(a)}(M/N)
    =
    -
    \sum_{k=0}^{2 P N}
    \psi_{2 P}^{(a)}(k) \,
    \mathrm{e}^{\pi \mathrm{i} \frac{M}{N} \frac{k^2}{2 P}} \,
    B_1
    \left(
      \frac{k}{2 \, P \, N}
    \right)
  \end{gather}
  Here 
  we assume $M$ and $N$ are relatively prime integers,
  and $N>0$.
  We use $B_k(x)$ as the $k$-th Bernoulli polynomial
  \begin{equation*}
    \sum_{k=0}^\infty B_k(x) \, \frac{t^k}{k!}
    =
    \frac{t \, \mathrm{e}^{x t}}{\mathrm{e}^t -1 }
  \end{equation*}
  and we have
  \begin{align*}
    B_1(x) & = x - \frac{1}{2} 
    \\[2mm]
    B_2(x) &= x^2 - x + \frac{1}{6}
  \end{align*}
\end{prop}

\begin{proof}
  It is a standard result of applying the Mellin transformation,
  so we omit the proof.
  See Refs.~\citen{LawrZagi99a,DZagie01a}
  (also Refs.~\citen{KHikami03a,KHikami03c,KHikami04b}).
\end{proof}

We note that a limiting value of the Eichler integrals at
$\tau\to N\in\mathbb{Z}$
is a little bit simplified;
\begin{gather}
%
  \widetilde{\Phi}_{\vv{p}}^{\vv{\ell}}(N)
  =
  - P \cdot
  C_{\vv{p}}(\vv{\ell}) \cdot
  \left(
    \mathbf{T}^{\vv{\ell}}
  \right)^N
  \label{phi_tilder_integer}
  \\[2mm]
%
%
  \widetilde{\Psi}_P^{(a)}(N)
  =
  \left(1 - \frac{a}{P}\right) \,
  \mathrm{e}^{\frac{a^2}{2 P } \pi \mathrm{i} N}
\end{gather}
where the $\mathbf{T}$-matrix is defined in eq.~\eqref{T_matrix}, and
we set
\begin{equation}
  \label{define_C_p}
  C_{\vv{p}} (\vv{\ell})
  =
  \sum_{n=1}^{2 P}
  \chi_{2 P}^{\vv{\ell}}(n) \,
  B_2
  \left(
    \frac{n}{2 \, P }
  \right) 
\end{equation}

\begin{prop}
  The Eichler integrals have 
  a nearly modular property.
  Especially
  asymptotic expansions in $N\to\infty$ are given by
  \begin{gather}
    \label{nearly_modular_Phi_tilde}
    \widetilde{\Phi}_{\vv{p}}^{\vv{\ell}}(1/N)
    +
    \left(
      \frac{N}{\mathrm{i}}
    \right)^{3/2} \,
    \sum_{\vv{\ell^\prime}}
    \mathbf{S}_{\vv{\ell}}^{\vv{\ell^\prime}}
    \,
    \widetilde{\Phi}_{\vv{p}}^{\vv{\ell^\prime}}(-N)
    \simeq
    \sum_{k=0}^\infty
    \frac{
      L(-2 \, k - 1 , \chi_{2 P}^{\vv{\ell}})
    }{
      k!
    } \,
    \left(
      \frac{
        \pi \, \mathrm{i}
      }{
        2 \, P \, N
      }
    \right)^k
    \\[2mm]
    \label{nearly_modular_Psi_tilde}
    \widetilde{\Psi}_P^{(a)}(1/N)
    +
    \sqrt{\frac{N}{ \mathrm{i}}} \,
    \sum_{b=1}^{P-1}
    \mathbf{M}_a^b \,
    \widetilde{\Psi}_P^{(b)}(-N)
    \simeq
    \sum_{k=0}^{\infty}
    \frac{
      L(-2 \, k , \psi_{2 P}^{(a)})
    }{
      k !
    } \,
    \left(
      \frac{\pi \, \mathrm{i}}{
        2 \, P \, N
      }
    \right)^k
  \end{gather}
  Here
  $N\in \mathbb{Z}$,
  and
  a sum of quadruples $\vv{\ell^\prime}$
  runs over $D$-dimensional space.
  We mean that
  $L(s,\chi)$ is the Dirichlet $L$-series.
\end{prop}

\begin{proof}
  We follow a proof given by Zagier for a case of
  weight $1/2$~\cite{DZagie01a}
  and  of weight $3/2$~\cite{LawrZagi99a},
  and we shall give an outline of the proof below. 
  See also Ref.~\citen{KHikami03a} for a proof of
  eq.~\eqref{nearly_modular_Psi_tilde}.

  We introduce other Eichler integrals defined by
  \begin{gather}
    \widehat{\Phi}_{\vv{p}}^{\vv{\ell}}(z)
    =
    -\sqrt{\frac{P \, \mathrm{i}}{2 \, \pi^2}} \,
    \int_{z^*}^\infty
    \frac{
      \Phi_{\vv{p}}^{\vv{\ell}}(\tau)
    }{
      \left( \tau - z \right)^{3/2}
    } \,
    \mathrm{d} \tau
    \\[2mm]
    \widehat{\Psi}_P^{(a)}(z)
    =
    \frac{1}{\sqrt{ 2 \, P \, \mathrm{i}}}
    \int_{z^*}^\infty
    \frac{
      \Psi_P^{(a)}(\tau)
    }{
      \sqrt{\tau - z}
    } \,
    \mathrm{d} \tau
  \end{gather}
  both of which are defined for $z$ in the lower half plane, $z\in
  \mathbb{H}^-$,
  and $*$ denotes a complex conjugate.
  We see that the modular properties  of
  $\Phi_{\vv{p}}^{\vv{\ell}}(\tau)$ and
  $\Psi_P^{(a)}(\tau)$, especially the modular
  ${S}$-transformation~\eqref{Phi_under_S} and~\eqref{Psi_under_S},
  lead
  \begin{gather}
    \label{modular_Phi_hat}
    \widehat{\Phi}_{\vv{p}}^{\vv{\ell}}(z)
    +
    \left(
      \frac{1}{  \   \mathrm{i} \, z   \  }
    \right)^{3/2} \,
    \sum_{\vv{\ell^\prime}}
    \mathbf{S}_{\vv{\ell}}^{\vv{\ell^\prime}}
    \,
    \widehat{\Phi}_{\vv{p}}^{\vv{\ell^\prime}}(-1/z)
    =
    r_{\Phi_{\vv{p}}}^{\vv{\ell}}(z; 0)
    \\[2mm]
    \label{modular_Psi_hat}
    \widehat{\Psi}_P^{(a)}(z)
    +
    \frac{  \  1 \   }{\sqrt{ \mathrm{i} \, z}} \,
    \sum_{b=1}^{P-1}
    \mathbf{M}_b^a \, \widehat{\Psi}_P^{(b)}(-1/z)
    =
    r_{\Psi_P}^{(a)}(z; 0)
  \end{gather}
  where
  \begin{gather*}
    r_{\Phi_{\vv{p}}}^{\vv{\ell}}(z; \alpha)
    =
    -\sqrt{\frac{P \, \mathrm{i}}{2 \, \pi^2}} \,
    \int_\alpha^\infty
    \frac{
      \Phi_{\vv{p}}^{\vv{\ell}}(\tau)
    }{
      \left( \tau - z \right)^{3/2}
    } \,
    \mathrm{d} \tau
    \\[2mm]
    r_{\Psi_P}^{(a)}(z; \alpha)
    =
    \frac{1}{\sqrt{ 2 \, P \, \mathrm{i}}}
    \,
    \int_{\alpha}^\infty
    \frac{
      \Psi_P^{(a)}(\tau)
    }{
      \sqrt{\tau - z}
    } \,
    \mathrm{d} \tau
  \end{gather*}
  with $\alpha\in\mathbb{Q}$.
  By substituting definitions of modular forms, we find that
  \begin{gather*}
    \widetilde{\Phi}_{\vv{p}}^{\vv{\ell}}(1/N)
    =
    \widehat{\Phi}_{\vv{p}}^{\vv{\ell}}(1/N)
    \\[2mm]
    \widetilde{\Psi}_P^{(a)}(1/N)
    =
    \widehat{\Psi}_P^{(a)}(1/N)
  \end{gather*}
  where LHSs are the limiting value from
  the upper plane $\mathbb{H}$ while RHSs are
  from the lower half plane
  $\mathbb{H}^-$.
  Taking asymptotic expansions of RHSs of
  both eqs.~\eqref{modular_Phi_hat}
  and~\eqref{modular_Psi_hat}, we obtain
  eqs.~\eqref{nearly_modular_Phi_tilde}
  and~\eqref{nearly_modular_Psi_tilde}.
\end{proof}

\section{Quantum Invariant and Eichler Integrals}
\label{sec:invariant_Eichler}

So far
we have studied both the WRT invariant for the Seifert manifold
and
the modular forms with half-integral weight, and have computed
limiting values of the Eichler integrals.
One of our main theorems is that the WRT invariant for the Seifert
manifold is written as a sum of the Eichler integrals.
So the quadruple $\vv{p}=(p_1, p_2, p_3, p_4)$ used to define the
modular form should be identified with the surgery data of the Seifert
manifold.

\begin{theorem}
  \label{theorem:WRT_Eichler}
  The WRT invariant for the Seifert homology sphere
  $\Sigma(p_1, p_2, p_3, p_4)$,
  which is computed in eq.~\eqref{Rozansky},
  is rewritten in
  terms of a limiting value of the Eichler integrals.
  \begin{itemize}
  \item       $\sum_{j=1}^4 \frac{1}{p_j} < 1 $;
    \begin{multline}
      \label{WRT_Eichler_1}
      \mathrm{e}^{
        \frac{2 \pi \mathrm{i}}{N}
        \left( \frac{\phi}{4} - \frac{1}{2} \right)
      } \,
      \left(
        \mathrm{e}^{\frac{2 \pi \mathrm{i}}{N}} - 1
      \right)
      \,    \tau_N \bigl( \Sigma(p_1,p_2,p_3, p_4) \bigr)
      \\
      =
      \frac{1}{4 \, P}\,
      \widetilde{\Phi}_{\vv{p}}^{(p_1 -1, 1,1,1)}(1/N)
      -
      \frac{1}{4 \, P}
      \sum_{a=1}^{P-1}
      a \,
      \chi_{2 P}^{(p_1-1,1,1,1)}(a) \,
      \widetilde{\Psi}_P^{(a)}(1/N)
    \end{multline}

  \item   $\sum_{j=1}^4 \frac{1}{p_j} > 1$;
    \begin{multline}
      \mathrm{e}^{
        \frac{2 \pi \mathrm{i}}{N}
        \left( \frac{\phi}{4} - \frac{1}{2} \right)
      } \,
      \left(
        \mathrm{e}^{\frac{2 \pi \mathrm{i}}{N}} - 1
      \right)
      \,    \tau_N \bigl( \Sigma(p_1,p_2,p_3, p_4) \bigr)
      \\
      =
      \frac{1}{4 \, P}\,
      \widetilde{\Phi}_{\vv{p}}^{(p_1 -1, 1,1,1)}(1/N)
      -
      \frac{1}{4 \, P}
      \sum_{a=1}^{P-1}
      a \,
      \chi_{2 P}^{(p_1-1,1,1,1)}(a) \,
      \widetilde{\Psi}_P^{(a)}(1/N)
      \\
      +
      \frac{1}{2} \,
      \widetilde{\Psi}_P^{(2 P-  \sum_j \frac{P}{p_j})}
      (1/N)
    \end{multline}
  \end{itemize}
\end{theorem}

To prove this theorem, we use the following formula;
\begin{lemma}
  For $N, k \in \mathbb{Z}_{>0}$ and $0 \leq k \leq N- 1 $, we have
  \begin{equation}
    \label{omega_identity}
    \sum_{n=1}^{N-1}
    \frac{
      \mathrm{e}^{ \frac{2 \pi \mathrm{i}}{N} (k+1) n}
    }{
      \left(
        1 - \mathrm{e}^{\frac{2 \pi \mathrm{i}}{N} n}
      \right)^2
    }
    =
    \frac{1}{12} -
    \frac{N^2}{2} \,
    B_2 \left(\frac{k}{N}\right)
  \end{equation}
\end{lemma}
\begin{proof}
  We set
  $\omega=\exp\left(\frac{2 \, \pi \, \mathrm{i}}{N}\right)$
  for brevity.
  We recall a trivial identity
  \begin{equation}
    \label{polynomial_trivial}
    (1-x) \, \sum_{c=1}^{N-1} c \, x^c
    =
    \sum_{c=1}^N x^c - N \, x^N 
  \end{equation}
  Substituting $x=\omega^a$ for the above with $a\in\mathbb{Z}$ satisfying
  $0< a <N$, we get
  \begin{equation}
    \frac{1}{1 - \omega^a}
    =
    -\frac{1}{N} \sum_{c=1}^{N-1} c \, \omega^{a c}
  \end{equation}
  where we have used $\sum_{c=1}^N \omega^{a c} = 0$.
  Differentiating eq.~\eqref{polynomial_trivial} w.r.t. $x$ and
  substituting
  $x=\omega^a$,
  we get
  \begin{equation}
    -2 \, N \, \frac{\omega^a}{\left( 1- \omega^a \right)^2}
    - N^2 \, \frac{1}{1- \omega^a}
    =
    \sum_{c=1}^{N-1} c^2 \, \omega^{a c}
  \end{equation}
  We then find
  \begin{equation*}
    \frac{\omega^a}{\left( 1 - \omega^a \right)^2}
    = \frac{1}{2 \, N}
    \sum_{c=0}^{N-1} c \, (N-c) \, \omega^{a c}
  \end{equation*}
  Using this expression, we see that
  \begin{align*}
    \text{LHS of \eqref{omega_identity}}
    & =
    \frac{1}{2} \sum_{m=1}^{N-1} \sum_{c=0}^{N-1}
    c \, \left(1 - \frac{c}{N}\right) \,
    \omega^{m (k+c)}
    \\
    & =
    \frac{1}{2} \sum_{c=0}^{N-1} c \, \left(1 - \frac{c}{N} \right) \,
    \left(
      -1 + N \, \left( \delta_{k+c,0} + \delta_{k+c,N} \right)
    \right)
    \\
    & =
    \frac{1}{12} \, \left(
      1 - N^2 + 6 \, N \, k - 6 \, k^2
    \right)
  \end{align*}
  which proves eq.~\eqref{omega_identity}.
  See  Ref.~\citen{ArakIbukKane01} for elegant treatments of
  several
  identities concerning  the $N$-th root of unity.
\end{proof}

\begin{proof}[Proof of Theorem~\ref{theorem:WRT_Eichler}]
  We first consider  a case  $\sum_j \frac{1}{p_j} < 1$.
  We find that there is a generating function for the periodic
  function
  $\chi_{2 P}^{(1,1,1,1)}(n)$;
  \begin{equation}
    \label{product_chi}
    - z^P \prod_{j=1}^4 \left( z^{P/p_j} - z^{- P/p_j} \right)
    =
    \sum_{n=0}^{2 P - 1}
    \chi_{2 P}^{(1,1,1,1)}(n) \, z^n
  \end{equation}
  Using above identity, we get
  \begin{align*}
    & \text{LHS of eq.~\eqref{WRT_Eichler_1}}
    \\
    & =
    \frac{ - \mathrm{e}^{\frac{\pi \mathrm{i}}{4}}}{
      2 \sqrt{2 \, P \, N}
    }
    \sum_{j=0}^{2 P -1 } \sum_{m=1}^{N-1} \sum_{k=0}^{2 P -1}
    \mathrm{e}^{
      - \frac{ ( N j + m - k)^2}{2 P N} \pi \mathrm{i}
      +
      \frac{ k^2}{2 P N} \pi \mathrm{i}
    } \,
    \chi_{2 P}^{(1,1,1,1)}(k) \,
    \frac{
      \mathrm{e}^{(j+\frac{m}{N}) \pi \mathrm{i}}
    }{
      \left(
        1 - \mathrm{e}^{\frac{2 \pi \mathrm{i}}{N} m}
      \right)^2
    }
    \\
    & =
    \frac{-1}{2 \, N}
    \sum_{k=0}^{2 P -1} \sum_{j=0}^{N-1}
    \chi_{2 P}^{(1,1,1,1)}(k) \,
    \mathrm{e}^{\frac{\pi \mathrm{i}}{2 P N}
      \left( k - 2 P ( j + \frac{1}{2}) \right)^2
    }
    \sum_{m=1}^{N-1}
    \frac{
      \mathrm{e}^{\frac{2 \pi \mathrm{i}}{N}(j+1) m}
    }{
      \left(
        1 - \mathrm{e}^{\frac{2 \pi \mathrm{i}}{N} m}
      \right)^2
    }
    \\
    & =
    -\frac{1}{2 \, N}
    \sum_{n=0}^{2 P N -1}
    \chi_{2 P}^{(1,1,1,1)}(n) \,
    \mathrm{e}^{
      \frac{\pi \mathrm{i}}{2 P N} ( n - P)^2
    } \,
    \left\{
      \frac{1}{12}
      -
      \frac{N^2}{2} \,
      B_2  \left( \frac{1}{N}
        \left\lfloor \frac{n}{2 \, P} \right\rfloor
      \right)
    \right\}
  \end{align*}
  Here in the second equality, we have used the Gauss sum reciprocity
  formula (see \emph{e.g.} Ref.~\citen{LCJeff92a})
  \begin{equation}
    \label{Gauss_sum_formula}
    \sum_{n \mod N}
    \mathrm{e}^{\frac{\pi \mathrm{i}}{N} M n^2 + 2 \pi \mathrm{i} k n}
    =
    \sqrt{
      \left|
        \frac{N}{M}
      \right|
    } \,
    \mathrm{e}^{\frac{\pi \mathrm{i}}{4} \sign(N M)}
    \sum_{n \mod M}
    \mathrm{e}^{
      -\frac{\pi \mathrm{i}}{M}
      N (n+k)^2
    }
  \end{equation}
  where $N, M\in \mathbb{Z}$
  with
  $N \, k \in \mathbb{Z}$ and
  $N \, M$ being even.
  In the third equality we have applied eq.~\eqref{omega_identity},
  and have used  properties of the Bernoulli polynomials,
  \begin{gather*}
    B_k(1-x) = (-1)^k \, B_k(x) 
    \\[2mm]
    B_k(x+1) - B_k(x) = k \, x^{k-1}
  \end{gather*}
  We then obtain
  \begin{multline}
    \label{middle_expression}
    \text{LHS of eq.~\eqref{WRT_Eichler_1}}
    =
    \frac{N}{4} \sum_{n=0}^{2 P N -1}
    \chi_{2 P}^{(1,1,1,1)}(n+P) \,
    \mathrm{e}^{\frac{n^2}{2 P N} \pi \mathrm{i}} 
    \\
    \times    
    \left\{
      B_2 \left( \frac{n}{2 \, P \, N} \right)
      -\frac{2}{N} \,
      B_1 \left( \frac{n}{2 \, P \, N} \right) \,
      B_1 \left( \frac{n+P}{2 \, P} -
        \left\lfloor \frac{n+P}{2 \, P} \right\rfloor
      \right)
      \right.
      \\
      \left.
      + \frac{1}{N^2} \,
      B_2 \left( \frac{n+P}{2 \, P} -
        \left\lfloor \frac{n+P}{2 \, P} \right\rfloor
      \right)
      -
      \frac{1}{12 \, N^2}
    \right\}
  \end{multline}
  Using  eq.~\eqref{chi_sigma},
  one sees that the first term gives
  $\frac{1}{4 \, P} \,
  \widetilde{\Phi}_{\vv{p}}^{\sigma_i(1,1,1,1)}(1/N)$,
  and that the second term is written in terms of
  $\widetilde{\Psi}_P^{(a)}(1/N)  $
  as the function
  $B_1
  \left(
    x+\frac{1}{2} - \left\lfloor x+\frac{1}{2}\right\rfloor
  \right)$ is an odd
  periodic   sawtooth function
  satisfying
  \begin{equation}
    \chi_{2 P}^{\vv{\ell}}(n) \,
    B_1 \left( \frac{n+P}{2 \, P} -
      \left\lfloor \frac{n+P}{2 \, P} \right\rfloor
    \right)
    =
    \frac{1}{2 \, P}
    \sum_{a=1}^P
    a \, 
    \chi_{2 P}^{\vv{\ell}}(a)
    \, \psi_{2 P}^{(a)}(n)
  \end{equation}

  To obtain eq.~\eqref{WRT_Eichler_1},
  we need to prove that the 
  remaining terms vanish.
  To see this,
  we have  
  for a case
  of the fourth  constant term in eq.~\eqref{middle_expression}
  \begin{align*}
    & \sum_{n=0}^{2 P N -1}
    \chi_{2 P}^{(1,1,1,1)}(n+P) \,
    \mathrm{e}^{\frac{n^2}{2 P N} \pi \mathrm{i}}
    \\
    & =
    -
    \frac{\mathrm{e}^{\frac{\pi \mathrm{i}}{4}}}{\sqrt{
        2 \, P \, N}} \,
    \sum_{n=0}^{2 P N-1}
    \sum_{k=0}^{2 P N - 1}
    \chi_{2 P}^{(p_1-1,1,1,1)}(n)
    \mathrm{e}^{- \frac{k^2}{2 P N} \pi \mathrm{i}
      + \frac{k n}{P N}\pi \mathrm{i}}
    \\
    & =
    -
    \frac{\mathrm{e}^{\frac{\pi \mathrm{i}}{4}}}{\sqrt{
        2 \, P \, N}} \,
    \sum_{j=0}^{2 P -1}
    \sum_{k=0}^{2 P N - 1}
    \chi_{2 P}^{(p_1-1,1,1,1)}(j)
    \,
    \left(
      \sum_{m=0}^{N-1}
      \mathrm{e}^{\frac{2 k m}{N} \pi \mathrm{i}}
    \right) \,
    \mathrm{e}^{- \frac{k^2}{2 P N} \pi \mathrm{i}
      + \frac{k j}{P N}\pi \mathrm{i}}
  \end{align*}
  where in the first equality we have used eq.~\eqref{chi_sigma} and
  eq.~\eqref{Gauss_sum_formula}.
  In the second equality we have simply  set $n= 2 \, P \, m + j$.
  As the sum in the parenthesis  in the last expression
  vanishes, we   can conclude that the
  last term in eq.~\eqref{middle_expression}
  vanishes.
  When we recall  the Fourier expansion of the periodic Bernoulli
  polynomials
  (cf. Ref.~\citen{ArakIbukKane01})
  \begin{equation*}
    B_k
    \left( x- \lfloor x \rfloor \right)
    = -
    \frac{k!}{\left(2 \, \pi \, \mathrm{i}\right)^k} \,
    \sum_{\substack{
        n \in \mathbb{Z}
        \\
        n \neq 0
      }}
    \frac{\mathrm{e}^{2 \pi \mathrm{i} n x}}{n^k}
  \end{equation*}
  for $k \in \mathbb{Z}_{>0}$,
  we can see
  that the third term in eq.~\eqref{middle_expression} also
  vanishes   in the same manner.
  This completes the  proof of Theorem~\ref{theorem:WRT_Eichler} for a
  case of $\sum_j \frac{1}{p_j} <1$.

  In  a case of $\sum_j \frac{1}{p_j} >1$, 
  the generating function~\eqref{product_chi} is
  replaced with
  \begin{multline}
    -z^P \, \prod_{j=1}^4
    \left( z^{P/p_j} - z^{- P/p_j}  \right)
    + z^P \, \left( z^P - z^{-P} \right) \,
    \left(
      z^{P \sum_j \frac{1}{p_j} - P}
      -
      z^{ -P \sum_j \frac{1}{p_j} + P}
    \right)
    \\
    =
    \sum_{n=0}^{2 P -1} \chi_{2 P}^{(1,1,1,1)}(n) \, z^n
    \label{product_chi_large}
  \end{multline}
  \emph{i.e.} we have an additional term coming from the second term
  in LHS of eq.~\eqref{product_chi_large}.
  Then 
  we have another contribution
  to the WRT invariant
  besides eq.~\eqref{middle_expression};
  \begin{equation}
    \frac{\mathrm{e}^{\frac{\pi \mathrm{i}}{4}}}{2
      \sqrt{2 \, P \, N}
    } \,
    \sum_{\substack{
        n= 0 \\
        N \, \nmid \, n
      }}^{2 \, P \, N-1 }
    \mathrm{e}^{-\frac{n^2}{2 P N} \pi \mathrm{i}} \,
    \frac{
      \mathrm{e}^{\frac{n}{N} \pi \mathrm{i}
        \left( \sum_j \frac{1}{p_j} - 1 \right)
      }
      -
      \mathrm{e}^{-\frac{n}{N} \pi \mathrm{i}
        \left( \sum_j \frac{1}{p_j} - 1 \right)
      }
    }{
      \mathrm{e}^{\frac{n}{N} \pi \mathrm{i}}
      -
      \mathrm{e}^{-\frac{n}{N} \pi \mathrm{i}}
    }
  \end{equation}
  This term can also be rewritten in terms of the Eichler integral.
  To see this fact, we compute as
  follows~\cite{LawrZagi99a,KHikami04b};
  \begin{align*}
    \widetilde{\Psi}_P^{(2 P - P\sum_j \frac{1}{p_j})}(1/N)
    & =
    \lim_{t \searrow 0}
    \sum_{n=0}^\infty
    \psi_{2 P}^{(2 P - P\sum_j \frac{1}{p_j})}(n)
    \,
    \mathrm{e}^{\frac{n^2}{2 P N} \pi \mathrm{i} - n t}
    \\
    & =
    \lim_{t \searrow 0}
    \sum_{n=0}^\infty
    \sum_{k=0}^{2 P N -1}
    \psi_{2 P}^{(2 P - P\sum_j \frac{1}{p_j})}(n)
    \frac{\mathrm{e}^{\frac{\pi \mathrm{i}}{4}}}{
      \sqrt{ 2 \, P \, N}
    } \,
    \mathrm{e}^{-\frac{k^2}{2 P N} \pi \mathrm{i}
      +\frac{k n}{P N} \pi \mathrm{i}
      - n t}
    \\
    & =
    \frac{\mathrm{e}^{\frac{\pi \mathrm{i}}{4}}}{
      \sqrt{2 \, P \, N}
    } \,
    \sum_{\substack{
        k=0 \\
        N \, \nmid \, k
      }}^{2 P N -1}
    \mathrm{e}^{-\frac{k^2}{2 P N} \pi \mathrm{i}} \,
    \frac{
      \mathrm{e}^{\frac{k}{N} \pi \mathrm{i}
        \left( \sum_j \frac{1}{p_j} - 1 \right)
      }
      -
      \mathrm{e}^{-\frac{k}{N} \pi \mathrm{i}
        \left( \sum_j \frac{1}{p_j} - 1 \right)
      }
    }{
      \mathrm{e}^{\frac{k}{N} \pi \mathrm{i}}
      -
      \mathrm{e}^{-\frac{k}{N} \pi \mathrm{i}}
    }
  \end{align*}
  In the last equality we have used a generating function
  of the periodic function
  $\psi_{2  P}^{(a)}(n)$, 
  and  also used a fact that the sum for $N | k$ vanishes.
  This completes the proof of the theorem.
\end{proof}

\section{Asymptotic Expansion of the WRT Invariant}
\label{sec:asymptotic}

We have  seen that the WRT invariant
$\tau_N(\mathcal{M})$
for the Seifert manifold
$\mathcal{M}=\Sigma(p_1, p_2, p_3, p_4)$ is written as a sum of the
Eichler
integrals of the modular forms with  half-integral weight.
As we have already
found   a nearly modular property of these Eichler integrals,
it is straightforward to obtain the following theorem.

\begin{theorem}
  \label{theorem:asymptotic_WRT}
  Asymptotic expansion of the WRT invariant
  $\tau_N(\mathcal{M})$
  for the Seifert homology
  sphere
  $\mathcal{M}=\Sigma(p_1, p_2, p_3, p_4)$
  in $N\to\infty$ is given as follows.
  \begin{itemize}
  \item       $\sum_{j=1}^4 \frac{1}{p_j} < 1 $;
    \begin{multline}
      \label{asymptotic_tau_1}
      \mathrm{e}^{
        \frac{2 \pi \mathrm{i}}{N}
        \left( \frac{\phi}{4} - \frac{1}{2} \right)
      } \,
      \left(
        \mathrm{e}^{\frac{2 \pi \mathrm{i}}{N}} - 1
      \right)
      \,    \tau_N \bigl( \Sigma(p_1,p_2,p_3, p_4) \bigr)
      \\
      \simeq
      -\frac{1}{4 \, P} \,
      \left(
        \frac{N}{\mathrm{i}}
      \right)^{3/2} \,
      \sum_{\vv{\ell}}
      \mathbf{S}_{p_1 -1, 1, 1,1}^{\vv{\ell}} \,
      \widetilde{\Phi}_{\vv{p}}^{\vv{\ell}}(-N)
      \\
      +
      \sqrt{\frac{N}{ \mathrm{i}}} \,
      \sum_{b=1}^{P-1}
      \left(
        \sum_{a=1}^{P-1} a \, \chi_{2 P}^{(p_1 -1,1,1,1)}(a) \,
        \sin \left(
          \frac{a \, b}{P} \, \pi
        \right)
      \right) \,
      \frac{P-b}{\sqrt{8 \, P^5}} \,
      \mathrm{e}^{- \frac{b^2}{2  \, P} \,  \pi \, \mathrm{i} \,  N}
      \\
      +
      \sum_{k=0}^\infty
      \frac{T_<(k)}{k !} \,
      \left(
        \frac{\pi \, \mathrm{i}}{2 \, P \, N}
      \right)^k
    \end{multline}
    where the $T$-series $T_<(k)$ is defined by
    \begin{align}
      T_<(k)
      & =
      \frac{1}{4 \, P} \,
      \left(
        L(-2 \, k -1 , \chi_{2 P}^{(p_1-1,1,1,1)})
        -
        \sum_{a=1}^{P-1}
        a \, \chi_{2 P}^{(p_1 - 1,1,1,1)}(a) \,
        L(-2 \, k , \psi_{2 P}^{(a)})
      \right)
      \label{define_T_small}
      \\
      & =
      \left( 2 \, P \right)^{2 k}
      \sum_{n= -P}^P
      \chi_{2 P}^{(p_1 -1, 1,1,1)}(n) \,
      \left(
        -\frac{1}{4 \, (k+1)} \,
        B_{2 k +2}\left(\frac{n}{2 \, P}\right)
        +
        \frac{n}{4 \, P \, (2 \, k+1)} \,
        B_{2 k+1} \left(\frac{n}{2 \, P}\right)
      \right)
      \nonumber
    \end{align}

    \item      $\sum_{j=1}^4 \frac{1}{p_j} > 1 $;
      \begin{multline}
        \label{asymptotic_tau_2}
        \mathrm{e}^{
          \frac{2 \pi \mathrm{i}}{N}
          \left( \frac{\phi}{4} - \frac{1}{2} \right)
        } \,
        \left(
          \mathrm{e}^{\frac{2 \pi \mathrm{i}}{N}} - 1
        \right)
        \,    \tau_N \bigl( \Sigma(p_1,p_2,p_3, p_4) \bigr)
        \\
        \simeq
        -\frac{1}{4 \, P} \,
        \left(
          \frac{N}{\mathrm{i}}
        \right)^{3/2} \,
        \sum_{\vv{\ell^\prime}}
        \mathbf{S}_{p_1 -1, 1, 1,1}^{\vv{\ell^\prime}} \,
        \widetilde{\Phi}_{\vv{p}}^{\vv{\ell^\prime}}(-N)
        \\
        +
        \sqrt{ \frac{N}{ \mathrm{i}}} \,
        \sum_{b=1}^{P-1}
        \left(
          \sum_{a=1}^{P-1} a \, \chi_{2 P}^{(p_1 -1,1,1,1)}(a) \,
          \sin \left(
            \frac{a \, b}{P} \, \pi
          \right)
        \right) \,
        \frac{P-b}{\sqrt{8 \, P^5}} \,
        \mathrm{e}^{-  \frac{b^2}{2 \,  P} \, \pi \, \mathrm{i} \, N}
        \\
        +
        \sqrt{ \frac{N}{ \mathrm{i} }} \,
        \sum_{b=1}^{P-1}
        \sin \left(
          \sum_j \frac{1}{p_j}  \, b \, \pi
        \right) \cdot
        \frac{P-b}{\sqrt{2 \, P^3}} \,
        \mathrm{e}^{-  \frac{b^2}{2 P} \, \pi \, \mathrm{i} \, N}
        \\
        +
        \sum_{k=0}^\infty
        \frac{T_>(k)}{k !} \,
        \left(
          \frac{\pi \, \mathrm{i}}{2 \, P \, N}
        \right)^k
      \end{multline}
  \end{itemize}
  where the $T$-series $T_>(k)$ is given by
  \begin{align}
    T_>(k)
    & =
    \frac{1}{4 \, P} \,
    \left(
      L(-2 \, k -1 , \chi_{2 P}^{(p_1-1,1,1,1)})
      -
      \sum_{a=1}^{P-1}
      a \, \chi_{2 P}^{(p_1 - 1,1,1,1)}(a) \,
      L(-2 \, k , \psi_{2 P}^{(a)})
    \right)
    \nonumber       \\
    & \qquad \qquad
    +
    \frac{1}{2 } \, 
    L(-2 \, k , \psi_{2 P}^{(2 P - P\sum_j \frac{1}{p_j})})
    \label{define_T_large}
    \\
    & =
    \left( 2 \, P \right)^{2 k}
    \sum_{n= -P}^P
    \chi_{2 P}^{(p_1 -1, 1,1,1)}(n) \,
    \left(
      -\frac{1}{4 \, (k+1)} \,
      B_{2 k +2}\left(\frac{n}{2 \, P}\right)
      +
      \frac{n}{4 \, P \, (2 \, k+1)} \,
      B_{2 k+1} \left(\frac{n}{2 \, P}\right)
    \right)
    \nonumber \\
    & \qquad \qquad
    +
    \frac{\left(2 \, P \right)^{2 k}}{2 \, k +1} \,
    B_{2k+1}
    \left(
      \frac{1}{2} \, \sum_j \frac{1}{p_j}
    \right)
    \nonumber
  \end{align}
\end{theorem}


\subsection{Lattice Points
  and Non-Vanishing Eichler Integral}

We consider dominating terms of the WRT
invariant~\eqref{asymptotic_tau_1}
and~\eqref{asymptotic_tau_2} in $N\to\infty$ in
detail.
We shall reveal a close connection with
the irreducible representation of the fundamental group
$\pi_1(\mathcal{M})$
and
the Chern--Simons invariant
$\CS(\mathcal{M})$
for the Seifert manifold
$\mathcal{M}=\Sigma(p_1, p_2, p_3, p_4)$.


Theorem~\ref{theorem:asymptotic_WRT} indicates that the WRT invariant has
exponentially divergent terms in $N\to\infty$.
Recalling an explicit form of the $\mathbf{S}$-matrix~\eqref{S_matrix}
we get the following formula.

\begin{coro}
  The Witten invariant for the Seifert homology sphere
  $\Sigma(p_1, p_2, p_3, p_4)$
  behaves in $N\to\infty$ as
  \begin{multline}
    \label{asymptotic_WRT_Seifert_4}
    Z_{N-2}\bigl(
    \Sigma( p_1, p_2, p_3, p_4)
    \bigr)
    \sim
    N \, \mathrm{e}^{\frac{3}{4} \pi \mathrm{i} - \frac{\phi}{2 N} \pi
      \mathrm{i}} \,
    \sum_{\vv{\ell}}
    (-1)^{1+ P \sum_k \frac{1}{p_k} + P \sum_{j \neq k}
      \frac{\ell_k}{p_j p_k}
    }
    \\
    \times
    \frac{2}{\sqrt{P}} \,
    \left(
      \prod_{k=1}^4 \sin
      \left(
        P \, \frac{\ell_k}{p_k^{~2} } \, \pi
      \right)
    \right) \,
    C_{\vv{p}}(\vv{\ell}) \,
    \mathrm{e}^{- \frac{P}{2} \left(
        1+\sum_k \frac{\ell_k}{p_k}
      \right)^2 \pi \mathrm{i} N
    }
  \end{multline}
  where the sum of
  $\vv{\ell}=(\ell_1, \ell_2, \ell_3, \ell_4)$ runs over
  $D$-dimensional space,
  and $C_{\vv{p}}(\vv{\ell})$ is defined in eq.~\eqref{define_C_p}.
\end{coro}

We note that,
when we collect all the exponentially divergent terms in
eq.~\eqref{asymptotic_tau_1} and~\eqref{asymptotic_tau_2}, we have 
\begin{equation}
  \label{WRT_with_next_term}
  Z_{N-2}\bigl(
  \Sigma( \vv{p})
  \bigr)
  \sim
  N \cdot Z_{N-2}^{(0)}
  \bigl(
  \Sigma(\vv{p})
  \bigr)
  + Z_{N-2}^{(1)}
  \bigl(
  \Sigma(\vv{p})
  \bigr)
\end{equation}
Here the leading term  $N \cdot Z_{N-2}^{(0)}(\Sigma(\vv{p}))$
comes from a limiting value~\eqref{phi_tilder_integer}
of the Eichler integral
$\widetilde{\Phi}_{\vv{p}}^{\vv{\ell}}(-N)$,
\begin{multline}
  Z_{N-2}^{(0)}
  \bigl(
  \Sigma(p_1, p_2, p_3, p_4)
  \bigr)
  =
  \mathrm{e}^{\frac{3}{4} \pi \mathrm{i}
    - \frac{\phi}{2 N}  \pi \mathrm{i}}
  \sum_{\vv{\ell}}
  (-1)^{
    1+\sum_k \frac{P}{p_k} + P \sum_{j \neq k} \frac{\ell_k}{p_j p_k}
  }
  \\
  \times
  \frac{2}{\sqrt{P}} \,
  \left(
    \prod_k \sin \left(
      P \, \frac{\ell_k}{p_k^{~2}} \, \pi
    \right)
  \right) \,
  C_{\vv{p}}(\ell) \,
  \mathrm{e}^{-  \frac{P}{2} \left(
      1 + \sum_k \frac{\ell_k}{p_k}
    \right)^2 \pi \mathrm{i} N
  }
\end{multline}
The term $Z_{N-2}^{(1)}$ denotes
the next leading term,
which follows from the limiting value of the
Eichler integral $\widetilde{\Psi}_{P}^{(a)}(-N)$;
\begin{itemize}
\item  for
  $\sum_j \frac{1}{p_j} >1$
  \begin{multline}
    Z_{N-2}^{(1)}\bigl(
    \Sigma(p_1, p_2, p_3, p_4)
    \bigr)
    \\
    =
    \mathrm{e}^{
      -\frac{3}{4} \pi \mathrm{i} - \frac{\phi}{2 N} \pi  \mathrm{i}
    }
    \sum_{b=1}^{P-1}
    \frac{P - b}{2 \sqrt{P^3} }    \,
    \left(
      \sin \Biggl( \sum_j \frac{b}{p_j} \pi \Biggr)
      -4 \sum_j \frac{1}{p_j} \, \cos \left(\frac{b}{p_j} \, \pi \right)
      \,
      \prod_{k \neq j}
      \sin \left( \frac{b}{p_k} \, \pi \right)
    \right)
    \,
    \mathrm{e}^{- 
      \frac{b^2}{2 \, P} \, \pi \, \mathrm{i}  \, N
    }
    \label{next_Z_1}
  \end{multline}

\item
  for $\sum_j \frac{1}{p_j} < 1$
  \begin{multline}
    Z_{N-2}^{(1)}\bigl(
    \Sigma(p_1, p_2, p_3, p_4)
    \bigr)
    \\
    =
    \mathrm{e}^{
      -\frac{3}{4} \pi \mathrm{i} - \frac{\phi}{2 N} \pi  \mathrm{i}
    }
    \sum_{b=1}^{P-1}
    \frac{2 \, (b-P)}{ \sqrt{P^3} }    \,
    \left(
      \sum_j \frac{1}{p_j} \, \cos \left(\frac{b}{p_j} \, \pi \right)
      \,
      \prod_{k \neq j}
      \sin \left( \frac{b}{p_k} \, \pi \right)
    \right)
    \,
    \mathrm{e}^{-     \frac{b^2}{2 \, P} \, \pi \, \mathrm{i} \, N
    }
    \label{next_Z_2}
  \end{multline}
\end{itemize}

The sum of quadruples $\vv{\ell}$ in the leading term
$Z_{N-2}^{(0)}
\bigl(
\Sigma(\vv{p})
\bigr)$
runs over $D$-dimensional space, but as  in the case of
the Brieskorn homology sphere
$\Sigma(p_1,p_2,p_3)$ with 3-singular fibers~\cite{KHikami04b},
the function $C_{\vv{p}}(\vv{\ell})$ may vanish for some
quadruples $\vv{\ell}$.

\begin{prop}
  \label{prop:simplex}
  Let $\gamma(p_1,p_2,p_3,p_4)$
  be the number of the Eichler integrals
  $\widetilde{\Phi}_{\vv{p}}^{\vv{\ell}}(\tau)$ which  do not vanish
  in a limit $\tau\to N\in \mathbb{Z}$.
  Then $D - \gamma(p_1, p_2, p_3, p_4)$ coincides with the number of the
  integral lattice points
  $\vv{\ell}=
  (\ell_1, \ell_2, \ell_3, \ell_4) \in \mathbb{Z}^4$
  satisfying
  \begin{equation}
    \label{condition_hypersurface}
    \begin{aligned}[t]
      0<
      \frac{\ell_1}{p_1} +\frac{\ell_2}{p_2} +\frac{\ell_3}{p_3}
      +\frac{\ell_4}{p_4}
      < 1
    \end{aligned}
  \end{equation}
  where
  $0< \ell_j < p_j$.
  Namely the number of lattice points inside the integral
  simplex whose vertices are
  $(p_1, 0,0,0)$,
  $(0,  p_2, 0, 0)$,
  $(0, 0, p_3, 0)$,
  $(0, 0, 0, p_4)$,
  and the origin $(0,0,0,0)$.
\end{prop}

\begin{proof}
  Eq.~\eqref{phi_tilder_integer}
  indicates that the limiting value
  $\widetilde{\Phi}_{\vv{p}}^{\vv{\ell}}(N)$
  of the Eichler integral
  does not
  vanish
  if and only if
  \begin{equation}
    C_{\vv{p}}(\vv{\ell}) \neq 0
  \end{equation}

  We consider a condition for quadruples $\vv{\ell}$.
  As
  we have $0<\frac{\ell_j}{p_j}<1$ by definition,  
  we have
  $ 0 <  \sum_{j=1}^4 \frac{\ell_j}{p_j} < 4$,
  and
  a pairwise coprime condition of $p_j$ indicates
  $\sum_j \frac{\ell_j}{p_j} \not\in \mathbb{Z}$.

  We first assume $0<\sum_j \frac{\ell_j}{p_j} <1$.
  Here we use $\{ a, b, c , d \} = \{ 1,2,3,4\}$.
  {}From $0<\frac{\ell_j}{p_j}<1$
  we see that
  $0<1 + \frac{\ell_a}{p_a} + \frac{\ell_b}{p_b} + \frac{\ell_c}{p_c}
  - \frac{\ell_d}{p_d} <2 $.
  If we have
  $-1<1 + \frac{\ell_a}{p_a} + \frac{\ell_b}{p_b} - \frac{\ell_c}{p_c}
  - \frac{\ell_d}{p_d} <0 $,
  we have a contradiction $\frac{\ell_a}{p_a}+\frac{\ell_b}{p_b}<0$  by summing
  with the  assumption  $0<\sum_j \frac{\ell_j}{p_j} <1$.
  Then we see that
  $0<1 + \frac{\ell_a}{p_a} + \frac{\ell_b}{p_b} - \frac{\ell_c}{p_c}
  - \frac{\ell_d}{p_d} <2 $
  for arbitrary setting of $\{a,b,c,d\}$.
  Then  
  collecting non-zero 16 terms of $\chi_{2 P}^{\vv{\ell}}(n)$,
  we can check  by a direct computation that
  \begin{align*}
    C_{\vv{p}}(\vv{\ell})
    & =
    -
    \sum_{\varepsilon_1,
      \varepsilon_2,
      \varepsilon_3,
      \varepsilon_4
      = \pm 1}
    \left(
      \prod_{j=1}^4 \varepsilon_j
    \right) \,
    B_2
    \left(
      \frac{1}{2}
      \left(
        1+\sum_{j=1}^4 \varepsilon_j \, \frac{\ell_j}{p_j}
      \right)
    \right)
    \\
    & =0
  \end{align*}
  
  We next consider a case of $1<\sum_j \frac{\ell_j}{p_j} < 3$.
  If we have
  $-1<1 + \frac{\ell_a}{p_a} + \frac{\ell_b}{p_b} - \frac{\ell_c}{p_c}
  - \frac{\ell_d}{p_d} <0$
  and
  $-1<1 + \frac{\ell_a}{p_a} - \frac{\ell_b}{p_b} + \frac{\ell_c}{p_c}
  - \frac{\ell_d}{p_d} <0$
  for some $\{a,b,c,d\}$,
  we get $1+\frac{\ell_a}{p_a} - \frac{\ell_d}{p_d}<0$ which contradicts with
  $0<\frac{\ell_j}{p_j}<1$.
  Thus,
  as inequalities
  for $1+\sum_j \varepsilon_j \frac{\ell_j}{p_j}$ with
  $\# \{\varepsilon_j=1\}=2$,
  we have two possibilities;
  \begin{itemize}
  \item 
    $0<1 + \frac{\ell_a}{p_a} + \frac{\ell_b}{p_b} - \frac{\ell_c}{p_c}
    - \frac{\ell_d}{p_d} <2$ for every setting of $\{a,b,c,d\}$,

  \item
    for a unique setting of $\{a,b,c,d\}$ among 6
    we have
    $-1<1 + \frac{\ell_a}{p_a} + \frac{\ell_b}{p_b} - \frac{\ell_c}{p_c}
    - \frac{\ell_d}{p_d} <0$
    and
    $2<1 - \frac{\ell_a}{p_a} - \frac{\ell_b}{p_b} + \frac{\ell_c}{p_c}
    + \frac{\ell_d}{p_d} <3$
  \end{itemize}
  In the latter case, we see easily
  $C_{\vv{p}}(\vv{\ell})=0$
  as the condition coincides with
  $0 < \sigma_{c d} \left(
    \sum_j \frac{\ell_j}{p_j}
  \right) <1$.
  For a case of the former,
  we still need to classify a condition for
  $1 + \frac{\ell_a}{p_a} + \frac{\ell_b}{p_b} + \frac{\ell_c}{p_c}
  - \frac{\ell_d}{p_d}$
  into
  $0<1 + \frac{\ell_a}{p_a} + \frac{\ell_b}{p_b} + \frac{\ell_c}{p_c}
  - \frac{\ell_d}{p_d}<2$
  or
  $2<1 + \frac{\ell_a}{p_a} + \frac{\ell_b}{p_b} + \frac{\ell_c}{p_c}
  - \frac{\ell_d}{p_d}<3$.
  \begin{itemize}
  \item 
    When
    $0<1 + \frac{\ell_a}{p_a} + \frac{\ell_b}{p_b} + \frac{\ell_c}{p_c}
    - \frac{\ell_d}{p_d}<2$
    for any setting of $\{a,b,c,d\}$, we have
    \begin{equation*}
      C_{\vv{p}}(\vv{\ell})
      =
      2 \, \left( \sum_j \frac{\ell_j}{p_j} -1 \right)
    \end{equation*}

  \item
    When, among $1+\sum_j \varepsilon_j \, \frac{\ell_j}{p_j}$ with
    $\# \{\varepsilon_j =-1\}=1$,
    we have, say
    $2<1 - \frac{\ell_a}{p_a} + \frac{\ell_b}{p_b} + \frac{\ell_c}{p_c}
    + \frac{\ell_d}{p_d}<3$ for a unique setting of $\{a,b,c,d\}$
    and
    others in $[0,2]$,
    we see
    \begin{equation*}
      C_{\vv{p}}(\vv{\ell})
      =
      4 \, \frac{\ell_a}{p_a}
    \end{equation*}

  \item
    When  two combinations of
    $1+\sum_j \varepsilon_j \, \frac{\ell_j}{p_j}$
    with $\#\{\varepsilon_j=-1\}=1$ are in $[2,3]$,
    say
    $2<1 - \frac{\ell_a}{p_a} + \frac{\ell_b}{p_b} + \frac{\ell_c}{p_c}
    + \frac{\ell_d}{p_d}<3$,
    $2<1 + \frac{\ell_a}{p_a} - \frac{\ell_b}{p_b} + \frac{\ell_c}{p_c}
    + \frac{\ell_d}{p_d}<3$,
    and others in $[0,2]$,
    we have
    \begin{equation*}
      C_{\vv{p}}(\vv{\ell})
      =
      2 \, \left(
        \frac{\ell_a}{p_a} + \frac{\ell_b}{p_b}
        -      \frac{\ell_c}{p_c} - \frac{\ell_d}{p_d}
      \right)
    \end{equation*}

  \item
    If  three combinations are in $[2,3]$,
    say
    $0<1 + \frac{\ell_a}{p_a} + \frac{\ell_b}{p_b} + \frac{\ell_c}{p_c}
    - \frac{\ell_d}{p_d}<2$
    for a unique setting of $\{a,b,c,d\}$,
    we get
    \begin{equation*}
      C_{\vv{p}}(\vv{\ell})
      =
      4 \, \left(
        1 - \frac{\ell_d}{p_d}
      \right)
    \end{equation*}
  \end{itemize}
  
  As a condition $3<\sum_j\frac{\ell_j}{p_j}<4$ means
  $0< \sigma_{1 2} \circ \sigma_{3 4}
  \left(
    \sum_j \frac{\ell_j}{p_j}
  \right) <1$, we get $C_{\vv{p}}(\vv{\ell})=0$ for this case.
  
  Combining these observations, we find that
  $C_{\vv{p}}(\vv{\ell})=0$ if
  $0 < \sum_j \frac{\ell_j}{p_j} <1$ or
  $0 < \sigma_{a b} \left( \sum_j \frac{\ell_j}{p_j} \right) <1$ or
  $3 < \sum_j \frac{\ell_j}{p_j}<4$.
  Recalling the invariance of $\chi_{2 P}^{\vv{\ell}}(n)$
  under the involution $\sigma_{a b}$, we can conclude
  that
  $D - \gamma(p_1, p_2, p_3, p_4 )$
  coincides with the number of integral lattice
  points satisfying eq.~\eqref{condition_hypersurface}.
\end{proof}

Computation of the number of the integral lattice points inside polytopes
is an old but difficult problem.
We have simple but beautiful
Pick's formula in the case of the 2-dimensional integral
polygons.
In the three-dimensional case, the number of the lattice points inside the
integral tetrahedron was computed by Mordell using the Dedekind sum~\cite{Morde51}.
In our case of the 4-dimensional simplex,
a formula was  also given by
Mordell~\cite{Morde51}, and Prop.~\ref{prop:simplex} proves the
following theorem.

\begin{theorem}
  The number of the non-vanishing Eichler integral
  $\widetilde{\Phi}_{\vv{p}}^{\vv{\ell}}(\tau)$
  at $\tau \to N \in \mathbb{Z}$ is given by
  \begin{multline}
    \label{gamma_formula}
    \gamma(p_1, p_2, p_3, p_4 )
    =
    - \frac{3}{8} +  \frac{P}{12} -
    \frac{P}{24} \, 
    \sum_{j=1}^4 \frac{1+  \,p_j}{p_j^{~2} }
    -
    \frac{1}{24 \, P} \,
    \left(
      1 - \sum_{j=1}^4 p_j
    \right)
    +
    \frac{P}{24}
    \sum_{j \neq  k }^4 \frac{1}{p_j^{~2} \, p_k}
    \\
    +
    \frac{1}{2}
    \sum_{j=1}^4
    s\left( \frac{P}{p_j} , p_j \right)
    -
    \frac{1}{2}
    \sum_{j \neq k}^4
    s\left(
      \frac{P}{p_j \, p_k}, p_j
    \right)
  \end{multline}
  where 
  $s(b,a)$ is the Dedekind sum~\eqref{Dedekind_sum}.
\end{theorem}

As the number $\gamma(\vv{p})$ of the non-vanishing Eichler integrals
corresponds to that
of the dominating terms~\eqref{asymptotic_WRT_Seifert_4} of the
WRT invariant,
and it could be related to the Casson invariant.
The Casson invariant of the Seifert manifold
$\Sigma ( p_1, p_2, \dots, p_M)$
with $M$-singular fibers
is defined  na\"{i}vely as  the number of the nontrivial SU(2)
representations of
$\pi_1\bigl(\Sigma(p_1,\dots,p_M)\bigr)$~\cite{KWalk92a}
(see also Refs.~\citen{Atiyah88a,NSave99Book}).
It  is known to be written  explicitly
as~\cite{FukuMatsSaka90,NeumWahl90a}
\begin{multline}
%
%
  \lambda_C
  \bigl( \Sigma( p_1, p_2, \dots , p_M) \bigr)
  \\
  =
  -\frac{1}{8} +
  \frac{1}{24 \, P_M} \,
  \left(
    1 + \sum_{j=1}^M 
    \left(
      \frac{P_M}{p_j}
    \right)^2
    - (M- 2) \, P_M^{~2}
  \right)
  -\frac{1}{2}
  \sum_{j=1}^M s \left(\frac{P_M}{p_j},p_j \right)
\end{multline}
where we have used $P_M=\prod_{j=1}^M p_j$.
Using this result  we obtain the following expression.
\begin{coro}
  We have
  \begin{multline}
    \label{explicit_gamma}
    \gamma(p_1, p_2, p_3, p_4)
    \\
    =
    \lambda_C \bigl( \Sigma(p_1, p_2, p_3) \bigr) +
    \lambda_C \bigl( \Sigma(p_1, p_2, p_4)\bigr)
    +  \lambda_C \bigl(\Sigma(p_1, p_3, p_4) \bigr)
    +  \lambda_C \bigl(\Sigma(p_2, p_3, p_4) \bigr)
    \\
    - \lambda_C \bigl( \Sigma(p_1, p_2, p_3, p_4) \bigr)
  \end{multline}
  where $\lambda_C(\mathcal{M})$ denotes the Casson invariant for
  3-manifold $\mathcal{M}$.
\end{coro}

We should note that,
in terms of the function $\phi(p_1,p_2,p_3,p_4)$
defined in eq.~\eqref{phi_definition}, we have~\cite{LawreRozan99a}
\begin{equation}
  -24 \cdot
  \lambda_C\bigl( \Sigma(p_1,p_2, p_3, p_4) \bigr)
  =
  \phi
  +
  P \,
  \left(
    2 - \sum_{j=1}^4 \frac{1}{p_j^{~2}}
  \right)
\end{equation}

The asymptotic behavior of the SU(2) WRT invariant for 3-manifold
$\mathcal{M}$
in $k\to\infty$
is expected to be~\cite{FreeGomp91a,EWitt89a}
\begin{equation}
  \label{expected_formula}
  Z_k( \mathcal{M} )
  \sim
  \frac{1}{2} \, \mathrm{e}^{- \frac{3}{4} \, \pi \,  \mathrm{i}} \,
  \sum_\alpha
  \sqrt{T_\alpha(\mathcal{M})} \,
  \mathrm{e}^{-2  \, \pi \, \mathrm{i} \,  I_\alpha/4}
  \,
  \mathrm{e}^{2 \pi \mathrm{i} (k+2) \CS(A)}
\end{equation}
where $T_\alpha(\mathcal{M})$ and $I_\alpha$ are respectively the
Reidemeister--Ray--Singer
torsion and the spectral flow,
and
the sum of  $\alpha$ denotes a
flat connection.

It is known~\cite{KHikami04b,Rozan95a} that,
in the case of the Brieskorn homology
sphere $\Sigma(p_1, p_2, p_3)$,
the exact asymptotic expansion of the WRT invariant has a form of
eq.~\eqref{expected_formula}
by identifying SU(2) representation of the fundamental group
$\pi_1(\mathcal{M})$
with flat connections on $\mathcal{M}$,
and that the Casson invariant
$\lambda_C\bigl(\Sigma(p_1,p_2,p_3)\bigr)$ 
is equal to
a minus one-half of the number of non-zero terms in
eq.~\eqref{expected_formula}.
Unlike the case of  the Brieskorn homology sphere,
we have seen that the
exact asymptotic expansion of the WRT invariant for
the Seifert manifold with 4-singular fibers
does not have a form of eq.~\eqref{expected_formula}
rather eq.~\eqref{asymptotic_WRT_Seifert_4}
as was pointed out in Ref.~\citen{Rozan95a}.
Especially the
number
of the dominating exponential terms is not proportional to the Casson
invariant,
and all the irreducible SU(2) representation of the fundamental
group~\eqref{fundamental_group} do not appear as 
we can read off from
eq.~\eqref{explicit_gamma}.

The
representation space
of the fundamental group
$\pi_1(\mathcal{M})$ of the Seifert manifold
$\mathcal{M}=\Sigma(p_1,p_2,p_3,p_4)$
with 4-singular fibers
was investigated in detail  in Refs.~\citen{KirkKlas91a,FintuStern90a}
(see also Ref.~\citen{NSave99Book}).
There are two types of the irreducible representation of
the fundamental group~\eqref{fundamental_group},
$\rho: \pi_1\bigl(\Sigma(p_1,p_2,p_3,p_4)\bigr)
\to SU(2)$,
up to conjugation;
\begin{itemize}

\item one of the generators $x_k$ is mapped to $\pm \id$,

\item all images $\rho(x_k)$ differ from $\pm \id$.

\end{itemize}
As  the former case can be given from
the representation space of the
Brieskorn homology sphere,
eq.~\eqref{explicit_gamma} shows that the number of lattice points
$\gamma(p_1,p_2,p_3,p_4)$ is related to the number of the latter case.
See Section~\ref{sec:example} for some examples.
We can thus conclude that the latter type of the irreducible
representations of $\pi_1(\mathcal{M})$ dominates the asymptotic
behavior of the WRT invariant in $N\to\infty$.

One may expect that the ``missing'' irreducible representations of $\pi_1(\mathcal{M})$,
in which one of  generators $x_k$ 
is mapped to $\pm \id$,
correspond to the next leading terms $Z_{N-2}^{(1)}(\mathcal{M})$~\eqref{next_Z_1} or~\eqref{next_Z_2}.
Actually they give the non-zero contribution, but
we can see that
the number of non-zero terms in $Z_{N-2}^{(1)}(\mathcal{M})$
is not equal to that of missing representations.

Once we  have seen that the dominating exponential factor of the WRT
invariant can be interpreted
from the SU(2) irreducible representation of the fundamental group
$\pi_1(\mathcal{M})$,
we can find that
the asymptotic behavior gives the Chern--Simons invariant of the manifold.
Explicitly
the  Chern--Simons invariant for the Seifert manifold is written
from
the exponential factor of eq.~\eqref{asymptotic_WRT_Seifert_4} as
\begin{equation}
  \label{CS_Seifert}
  \CS(A)
  = - \frac{P}{4} \,
  \left(
    1+\sum_{j=1}^4 \frac{\ell_j}{p_j}
  \right)^2
  \mod 1
\end{equation}
which originally appears as the
phase of the  $\mathbf{T}$-matrix~\eqref{T_matrix}
of the modular form 
$\Phi_{\vv{p}}^{\vv{\ell}}(\tau)$.

\subsection{Contribution from Trivial Connections}
In asymptotic behavior of the WRT invariant
$\tau_N(\mathcal{M})$
in $N\to\infty$, the exponential terms are 
dominating, but
a tail part has its own meaning as a contribution from trivial connections
(see  Refs.~\citen{LawreRozan99a,Rozan95a}).
We further
show that this corresponds to the Ohtsuki invariant for the Seifert manifold.

Before discussing a connection with the quantum invariant,
we give a generating function of the $T$-series in
eqs.~\eqref{define_T_small} and~\eqref{define_T_large}.

\begin{prop}
  Let the $T$-series $T_{\lessgtr}(k)$ be defined by
  eqs.~\eqref{define_T_small} and~\eqref{define_T_large}.
  We have
  \begin{equation}
    \label{coefficient_sinh}
    \frac{
      \prod_{j=1}^4 \sinh \left( \frac{P}{p_j} \, x \right)
    }{
      \bigl( \sinh ( P \, x ) \bigr)^2
    }
    =
    \frac{1}{2} \,
    \sum_{k=0}^\infty
    \frac{T_{\lessgtr}(k)}{( 2 \, k)!} \,
    x^{2 k}
  \end{equation} 
  depending on $\sum_{j=1}^4 \frac{1}{p_j} \lessgtr 1$.
\end{prop}

\begin{proof}
  We first study a case of $\sum_j \frac{1}{p_j} < 1$.
  We have  from eq.~\eqref{product_chi}
  \begin{align*}
    \frac{
      \prod_{j=1}^4 \left( z^{P/p_j} - z^{-P/p_j} \right)
    }{
      \left( z^P - z^{-P} \right)^2
    }
    & =
    -\sum_{n=0}^\infty
    \left(
      1+
      \left\lfloor
        \frac{n}{2 \, P}
      \right\rfloor
    \right) \,
    \chi_{2 P}^{(1,1,1,1)}(n) \, z^{n+P}
    \\
    & =
    \sum_{k=0}^\infty \sum_{n=P}^{3 P -1}
    (1+k) \, \chi_{2 P}^{(p_1 -1, 1,1,1)}(n) \, z^{2 P k + n}
  \end{align*}
  We substitute $z=\mathrm{e}^{-x}$ for the  above expression, and we
  equate it with
  $
  \sum_{k=0}^\infty T_{k} \, x^{2 k}
  $ in a limit $x\searrow 0 $.
  Applying the Mellin transformation,
  we get
  \begin{equation}
    \label{gamma_zeta}
    T_k
    =
    \frac{(2 \, P)^{2 k} }{(2 \, k)!}
    \sum_{n=P}^{3 P -1 }
    \chi_{2 P}^{(p_1-1,1,1,1)}(n) \,
    \left(
      \zeta \left(- 2 \, k-1,\frac{n}{2\, P}\right)
      -
      \frac{n-2 \, P}{2 \, P} \,
      \zeta \left( -2 \, k, \frac{n}{2 \, P}\right)
    \right)
  \end{equation}
  Using an analytic continuation
  of the Hurwitz zeta function
  \begin{equation*}
    \zeta( 1-k, z ) = - \frac{B_k(z)}{k}
  \end{equation*}
  for $k \in \mathbb{Z}_{>0}$,
  we get the  statement of the theorem.

  In  a case of $\sum_j \frac{1}{p_j} >  1$, we recall
  eq.~\eqref{product_chi_large} which gives
  \begin{equation*}
    \frac{
      \prod_{j=1}^4 \left( z^{P/p_j} - z^{-P/p_j} \right)
    }{
      \left( z^P - z^{-P} \right)^2
    }
    =
    \sum_{k=0}^\infty \sum_{n=P}^{3 P -1}
    (1+k) \, \chi_{2 P}^{(p_1 -1, 1,1,1)}(n) \, z^{2 P k + n}
    +
    \sum_{k=0}^\infty
    \psi_{2 P}^{(2 P - P \sum_j \frac{1}{p_j})}(k) \,
    z^k
  \end{equation*}
  By the Mellin transformation after setting $z=\mathrm{e}^{-x}$,
  the first term of RHS gives eq.~\eqref{gamma_zeta}
  while
  the second term gives
  $\frac{1}{(2 k)!} \,
  L(-2 \, k, \psi_{2 P}^{(2 P - P \sum_j \frac{1}{p_j})})$
  as a  coefficient of $x^{2 k}$.
  Combining these results, we obtain eq.~\eqref{coefficient_sinh}.
\end{proof}

We note that some of the $T$-series are explicitly computed as follows;
\begin{align*}
  T_\lessgtr(0) & = 0
  \\[2mm]
  T_\lessgtr(1) & =  4 \, P
  \\[2mm]
  T_\lessgtr(2) & =  8 \, P^3 \,
  \left( -2 + \sum_{j=1}^4 \frac{1}{p_j^{~2}} \right)
  \\[2mm]
  T_\lessgtr(3) &=
  4 \, P^5 \,
  \left(
    5 \,
    \Bigl(
      2 - \sum_{j=1}^4 \frac{1}{p_j^{~2}}
    \Bigr)^2
    +2 \,
    \Bigl(
      2-\sum_{j=1}^4 \frac{1}{p_j^{~4}}
    \Bigr)
  \right)
\end{align*}

The fact that coefficients in
tail part  $T_\lessgtr(k)$ of the asymptotic expansion of the WRT
invariant
also appear in the Taylor series~\eqref{coefficient_sinh} was pointed out
in Ref.~\citen{LawreRozan99a} by the different manner.
It is noted that
we also have a similar connection between the Eichler integral and the
colored Jones polynomial for torus knots/links, and that
the inverse of the Alexander polynomial generates this
tail part of the $N$-colored
Jones polynomial at the $N$-th root of unity.

See that the generating function~\eqref{coefficient_sinh} of
$T_{\lessgtr}(k)$ has already appeared in the WRT
invariant~\eqref{Rozansky} of the Seifert manifold.
The $T$-series is still related to the quantum invariant, \emph{i.e.},
the Ohtsuki invariant.
The $n$-th Ohtsuki invariant $\lambda_n(\mathcal{M})$ for 3-manifold
$\mathcal{M}$
is defined by~\cite{HMuraka93a,TOhtsu96a}
\begin{align}
  \tau_\infty
  (
  \mathcal{M}
  )
  =
  \sum_{n=0}^\infty \lambda_n (  \mathcal{M} ) \,
  (q-1)^n
\end{align}
where  the formal power series $\tau_\infty(\mathcal{M})$
is defined  from a tail part of the asymptotic expansion of the WRT
invariant,
and in our  case of the Seifert manifold 
$\mathcal{M}=\Sigma(p_1,p_2,p_3,p_4)$
with 4-singular fibers
we have from Theorem~\ref{theorem:asymptotic_WRT} that
\begin{equation}
  q^{\frac{\phi}{4} - \frac{1}{2}} \,
  (q-1) \cdot
  \tau_\infty
  \bigl(
  \Sigma(p_1,p_2,p_3,p_4)
  \bigr)
  =
  \sum_{k=0}^\infty
  \frac{T_\lessgtr(k)}{k!}
  \,
  \left(
    \frac{\log q }{4 \, P}
  \right)^k
\end{equation}
if we replace
 the $N$-th root of unity with a parameter $q$,
$q
\leftrightarrow
\exp\left(\frac{2 \, \pi \, \mathrm{i}}{N}\right)$.
Using the Stirling number of the first kind $S_n^{(m)}$ defined by
\begin{equation*}
  \prod_{j=0}^{n-1} (x-j)
  =
  \sum_{m=0}^n S_n^{(m)} \, x^m
\end{equation*}
which satisfies
(see \emph{e.g.} Ref.~\citen{ArakIbukKane01})
\begin{equation*}
  \frac{
    \left( \log q \right)^m
  }{m!}
  =
  \sum_{n=m}^\infty S_n^{(m)} \,
  \frac{\left(q - 1\right)^n}{n!}
\end{equation*}
we obtain the following formula;
\begin{prop}
  The $n$-th Ohtsuki invariant $\lambda_n(\mathcal{M})$
  for the Seifert manifold
  $\mathcal{M}=\Sigma(p_1,p_2,p_3,p_4)$
  is given by
  \begin{equation}
    \lambda_n \bigl(
    \Sigma(p_1, p_2, p_3, p_4)
    \bigr)
    =
    \sum_{j=0}^n
    \begin{pmatrix}
      \frac{2-\phi}{4} \\
      n-j
    \end{pmatrix}
    \sum_{k=1}^{j+1}
    \frac{T_\lessgtr(k)}{(4 \, P)^k} \,
    \frac{S_{j+1}^{(k)}}{(j+1)!}
  \end{equation}
\end{prop}

One can check that
the first three terms
are computed explicitly as follows;
\begin{align*}
  \lambda_0 & = 1 
  \\[2mm]
  \lambda_1 & =
  6 \, \lambda_C \bigl( \Sigma(p_1,p_2, p_3, p_4) \bigr)
  \\[2mm]
  \lambda_2 & =
  \frac{3 \, \phi^2 + 12 \, \phi + 4}{96}
  +
  \frac{P}{16} \, (\phi+2) \,
  \Bigl(
    2- \sum_j \frac{1}{p_j^{~2}}
  \Bigr)
  \\
  & \qquad
  +
  \frac{P^2}{96} \,
  \left(
    2 \, \Bigl( 2 - \sum_j \frac{1}{p_j^{~4}} \Bigr)
    +
    5 \, \Bigl( 2 - \sum_j \frac{1}{p_j^{~2}} \Bigr)^2
  \right)
\end{align*}
The fact that $\lambda_1$ is equal to 6 times of the Casson invariant
was
first pointed out in
Ref.~\citen{HMuraka93a},
and  $\lambda_2$ was derived in Ref.~\citen{CSato97a}.
See also Ref.~\citen{RLawren99a}.

\section{Examples}
\label{sec:example}
\subsection{$\Sigma(2,3,5,7)$}

With $\vv{p}=(2,3,5,7)$
the modular form spans $D=6$ dimensional space, and the
independent periodic functions $\chi_{420}^{\vv{\ell}}(n)$
are defined when quadruples
$\vv{\ell}$
are
$(1,1,1,1)$,
$(1,1,1,2)$,
$(1,1,1,3)$,
$(1,1,2,1)$,
$(1,1,2,2)$,
and
$(1,1,2,3)$.
We have $\sum_j \frac{1}{p_j} = \frac{247}{210}>1$, and
we can check $C_{\vv{p}}(\vv{\ell})\neq 0$ for all $\vv{\ell}$.
Indeed from eq.~\eqref{gamma_formula} we have
$\gamma(2,3,5,7)=6$.
These quadruples correspond to the
irreducible SU(2) representation of the
fundamental group~\eqref{fundamental_group},
in which none of the generators $x_k$ is  mapped
to $\pm \id$.
The Casson invariant is computed as
$\lambda_C\bigl(\Sigma(2,3,5,7)\bigr)=-14$, and
we have missing
irreducible representations in which one of generators is mapped to
$\pm \id$.

See Tables~\ref{tab:2357} and~\ref{tab:2357_2},
which should be compared with a table in Ref.~\citen{FintuStern90a}.
In Table~\ref{tab:2357}, collected are quadruples $\vv{\ell}$,
which contribute to the asymptotic behavior~\eqref{asymptotic_WRT_Seifert_4}
of the WRT invariant.
These denote the SU(2) representations
of $\pi_1(\mathcal{M})$, which do not 
map any generators $x_k$~\eqref{fundamental_group} to $\pm\id$.
We have listed quadruples $\vv{\ell}$ in
Table~\ref{tab:2357_2},
and they correspond to the irreducible
representations missing in Table~\ref{tab:2357}.
The number of these missing representations is proportional to the
sum of the 
Casson invariant,
$\lambda_C\bigl( \Sigma(2,3,5) \bigr)+
\lambda_C\bigl( \Sigma(2,3,7) \bigr)+
\lambda_C\bigl( \Sigma(2,5,7) \bigr)+
\lambda_C\bigl( \Sigma(3,5,7) \bigr)$.
As the
representation in Table~\ref{tab:2357} have 2-dimensional
components, 
we indeed recover the Casson invariant by
$ -\frac{1}{2}
\left(
  2 \times 6 + (2+2+4+8)
\right)=-14$.

\begin{table}[htbp]
  \centering
  \begin{equation*}
    \def\arraystretch{1.3}
    \begin{array}{c||ccc}
      \vv{\ell}
      & \sum_{j=1}^4 \frac{\ell_j}{p_j}
      & C_{\vv{p}}(\vv{\ell})
      &
      \text{$\CS(\mathcal{M})$}
      \\
      \hline        \hline
      (1,1,1,1)
      & \frac{247}{210} & \frac{37}{105} & -\frac{529}{840}
      \\
      (1,1,1,2) 
      &\frac{277}{210}  & \frac{67}{105} & -\frac{289}{840}
      \\
      (1,1,1,3) 
      &\frac{307}{210} & \frac{4}{5} & -\frac{169}{840}
      \\
      (1,1,2,1) 
      & \frac{289}{210} & \frac{4}{7} & -\frac{361}{840}
      \\
      (1,1,2,2) 
      & \frac{319}{210} & \frac{109}{105} & -\frac{121}{840}
      \\
      (1,1,2,3) 
      & \frac{349}{210} &  \frac{139}{105} & -\frac{1}{840}
    \end{array}
  \end{equation*}
  \caption{
    $\mathcal{M}=\Sigma(2,3,5,7)$:
    Listed are quadruples
    $\protect\vv{\ell}$, which contribute to the
    asymptotics of the
    WRT invariant $Z_N(\mathcal{M})$.
    These quadruples 
    denote  the irreducible
    representations of $\pi_1(\mathcal{M})$~\eqref{fundamental_group},
    in which none of the
    generators
    $x_k$
    is mapped to $\pm \id$.
    The Chern--Simons invariant
    $\CS(\mathcal{M})$
    is computed from eq.~\eqref{CS_Seifert}.}
  \label{tab:2357}
\end{table}

\begin{table}[htbp]
  \centering
  \begin{equation*}
    \def\arraystretch{1.3}
    \begin{array}{c||cc}
      \vv{\ell}
      &
      \sum_j \frac{\ell_j}{p_j}
      &
      \CS(\mathcal{M})
      \\
      \hline        \hline
      (1,1,1,0)  
      & \frac{31}{30} & -\frac{7}{120}
      \\
      (1,1,2,0)  
      & \frac{37}{30} & -\frac{103}{120}
      \\
      \hline
      (1,1,0,2)  
      & \frac{47}{42} & -\frac{125}{168}
      \\
      (1,1,0,3)  
      & \frac{53}{42} & -\frac{101}{168}
      \\
      \hline
      (1,0,1,3)  
      & \frac{79}{70} & -\frac{243}{280}
      \\
      (1,0,2,1)  
      & \frac{73}{70} & -\frac{27}{280}
      \\
      (1,0,2,2)  
      & \frac{83}{70} & -\frac{227}{280}
      \\
      (1,0,2,3)  
      & \frac{93}{70} & -\frac{187}{280}
      \\
      \hline
      (0,1,1,4)  
      & \frac{116}{105} & -\frac{121}{210}
      \\
      (0,1,1,5)  
      & \frac{131}{105} & -\frac{23}{105}
      \\
      (0,1,1,6)  
      & \frac{146}{105} & -\frac{1}{210}
      \\
      (0,1,2,2)  
      & \frac{107}{105} & -\frac{2}{105}
      \\
      (0,1,2,3)  
      & \frac{122}{105} & -\frac{79}{210}
      \\
      (0,1,2,4)  
      & \frac{137}{105} & -\frac{92}{105}
      \\
      (0,1,2,5)  
      & \frac{152}{105} & -\frac{109}{210}
      \\
      (0,1,2,6)  
      & \frac{167}{105} & -\frac{32}{105}
      \\
      \hline
    \end{array}
  \end{equation*}
  \caption{
    $\mathcal{M}=\Sigma(2,3,5,7)$:
    Listed are quadruples, which correspond to the representations
    of the fundamental group $\pi_1(\mathcal{M})$
    in which
    one of generators $x_k$
    is mapped to $\pm \id$.
  }
  \label{tab:2357_2}
\end{table}

In   Table~\ref{tab:2357_asymptotic} we give
a result of numerical computations,
and
compare with the exact value~\eqref{Rozansky} of the WRT invariant
with the asymptotic formula~\eqref{asymptotic_WRT_Seifert_4}.
We have also given numerical values~\eqref{WRT_with_next_term}
including the next leading terms
$Z_{N-2}^{(1)}(\mathcal{M})$
to find a good agreement.
All these computations are performed on PARI/GP.
We stress that the representations $\vv{\ell}$ in Table~\ref{tab:2357}
dominate  the asymptotic behavior of the WRT invariant
in $N\to\infty$.

  \begin{table}[htbp]
    \centering
    \begin{equation*}
      \def\arraystretch{1.03}
      \begin{array}{r||cc}
        N & \text{exact result for $Z_N$} &
        \text{asymptotic formula
          $
          \begin{cases}
            (N+2) \cdot Z_N^{(0)} + Z_N^{(1)} 
            \\
            (N+2) \cdot Z_N^{(0)}
          \end{cases}
          $}
        \\
        \hline \hline
        10 & 
        0.739637+
        2.732051 \, \mathrm{i}
        &
        0.740798+
        2.732420 \, \mathrm{i}
        \\
        & &
        1.259843+
        2.437384 \, \mathrm{i}
        \\
        11 &
        0.979154-
        0.903934 \, \mathrm{i}
        &
        0.980292-
        0.903502 \, \mathrm{i}
        \\
        & &
        0.879253-
        0.968693 \, \mathrm{i}
        \\
        12 &
        -0.01437482+
        0.0 \, \mathrm{i}
        &
        -0.01326139
        +0.00048566 \, \mathrm{i}
        \\
        & &
        -0.00073383
        +0.00867459 \, \mathrm{i}
        \\
        13 &
        -0.04815996-
        0.13669 \, \mathrm{i}
        &
        -0.04707020-
        0.136157687 \, \mathrm{i}
        \\
        & &
        -0.00575862+
        0.007899715 \, \mathrm{i}
        \\
        14 &
        -1.864025-
        0.7606090 \, \mathrm{i}
        &
        -1.862959-
        0.7600360 \, \mathrm{i}
        \\
        & &
        -1.995763-
        1.0444076 \, \mathrm{i}
        \\
        100 &
        -4.8885515 +
        17.7770857 \, \mathrm{i}
        &
        -4.8884355 +
        17.7779834 \, \mathrm{i}
        \\
        & &
        -4.8802978 +
        18.1353532 \, \mathrm{i}
        \\
        101 & 
        22.22565 +
        7.134420 \, \mathrm{i}
        &
        22.22576 +
        7.135315 \, \mathrm{i}
        \\
        & &
        22.39565 +
        7.174075 \, \mathrm{i}
        \\
        102 &
        -0.6760584-
        6.611630 \, \mathrm{i}
        &
        -0.6759530-
        6.610738 \, \mathrm{i}
        \\
        & &
        -0.6140020-
        6.398063 \, \mathrm{i}
        \\
        103 &
        0.28253575+
        0.2563299 \, \mathrm{i}
        &
        0.28263601+
        0.25721904 \, \mathrm{i}
        \\
        & &
        0.00417075+
        0.06066500 \, \mathrm{i}
        \\
        104 &
        -0.1814129 -
        6.39840405 \, \mathrm{i}
        &
        -0.1813177 -
        6.39751779 \, \mathrm{i}
        \\
        & &
        -0.2590138-
        6.54593985 \, \mathrm{i}
        \\
        1000 &
        -86.3814448-
        52.1955841 \, \mathrm{i}
        &
        -86.3815556-
        52.1955281 \, \mathrm{i}
        \\
        & &
        -86.3595930-
        52.0529208 \, \mathrm{i}
        \\
        1001 &
        -32.1688750+
        226.931025 \, \mathrm{i}
        &
        -32.1689857+
        226.931025 \, \mathrm{i}
        \\
        & &
        -32.0565661+
        227.078420 \, \mathrm{i}
        \\
        1002 &
        112.342695+
        21.8373199 \, \mathrm{i}
        &
        112.342584+
        21.8373757 \, \mathrm{i}
        \\
        & &
        112.122297+
        21.6486285 \, \mathrm{i}
        \\
        1003 &
        0.7904096329+
        0.9244664554 \, \mathrm{i}
        &
        0.7902992554 +
        0.9245221724 \, \mathrm{i}
        \\
        & &
        0.6333213866 +
        1.0639142783 \, \mathrm{i}
        \\
        1004 &
        57.8951433+
        129.671829 \, \mathrm{i}
        &
        57.8950330+
        129.671885 \, \mathrm{i}
        \\
        & &
        58.1582795+
        130.079477 \, \mathrm{i}
        \\
        1005 &
        69.6412229-
        74.3079505 \, \mathrm{i}
        &
        69.6411128-
        74.3078950 \, \mathrm{i}
        \\
        & &
        69.5928674-
        74.2038809 \, \mathrm{i}
        \\
        10000 &
        527.74686902+
        862.13517540 \, \mathrm{i}
        &
        527.74686459+
        862.13517563 \, \mathrm{i}
        \\
        & &
        527.55843581+
        861.79480366 \, \mathrm{i}
        \\
        10001 &
        -6.393230664+
        1.730198170 \, \mathrm{i}
        &
        -6.393235091+
        1.730198403 \, \mathrm{i}
        \\
        & &
        -6.624993671+
        1.763443057 \, \mathrm{i}
        \\
        10002 &
        -301.5164629-
        551.4353628 \, \mathrm{i}
        &
        -301.5164673-
        551.4353625 \, \mathrm{i}
        \\
        & &
        -301.7624718-
        551.3640027 \, \mathrm{i}
        \\
        10003 &
        -10.84155396-
        1.9919898321 \, \mathrm{i}
        &
        -10.84155838-
        1.9919895988 \, \mathrm{i}
        \\
        & &
        -10.95621827-
        1.9107737795 \, \mathrm{i}
        \\
        10004 &
        -868.9478695-
        736.82135025 \, \mathrm{i}
        &
        -868.9478739-
        736.8213500 \, \mathrm{i}
        \\
        & &
        -869.2651992-
        736.9897060 \, \mathrm{i}
      \end{array}
    \end{equation*}
    \caption{Numerical values of the WRT
      invariant $Z_N(\mathcal{M})$ for $\mathcal{M}=\Sigma(2,3,5,7)$.
      We have used $P=210$ and $\phi=\frac{949}{210}$ in
      eq.~\eqref{Rozansky}.
    }
    \label{tab:2357_asymptotic}
  \end{table}

  

\subsection{$\Sigma(3,4,5,7)$}

We have $\sum_j 1/p_j = \frac{389}{420}<1$, and
eq.~\eqref{define_D} gives
$D=18$.
We find that
$\gamma(3,4,5,7)=17$
from eq.~\eqref{gamma_formula}.
Indeed
among 18 independent quadruples $\vv{\ell}$, we see that
$C_{3,4,5,7}(1,1,1,1)=0$.
See Table~\ref{tab:3457} for representations $\vv{\ell}$,
in which none of generators $x_k$ of the fundamental group~\eqref{fundamental_group} is mapped
to $\pm \id$.
The Casson  invariant is computed as
$\lambda_C\bigl(\Sigma(3,4,5,7)\bigr)=-31$,
and we have missing  representations.
As the representations in Table~\ref{tab:3457} have 2-dimensional
components,
missing  representations can be given for the Seifert manifold with
3-singular fibers,
and its number is proportional to the sum of the Casson invariant,
$\lambda_C\bigl(\Sigma(3,4,5)\bigr)=-2$,
$\lambda_C\bigl(\Sigma(3,4,7)\bigr)=-3$,
$\lambda_C\bigl(\Sigma(3,5,7)\bigr)=-4$,
and
$\lambda_C\bigl(\Sigma(4,5,7)\bigr)=-5$.

\begin{table}[htbp]
  \centering
    \begin{equation*}
      \def\arraystretch{1.3}
      \begin{array}{c||ccc}
        \vv{\ell} 
        & \sum_j \frac{\ell_j}{p_j}
        & C_{\vv{p}}(\vv{\ell})
        &
        \CS(\mathcal{M})
        \\
        \hline        \hline
        (1,1,1,2) 
        & \frac{449}{420} & \frac{29}{210} & -\frac{841}{1680}
        \\
        (1,1,1,3) 
        & \frac{509}{420} & \frac{89}{210} & -\frac{1201}{1680}
        \\
        (1,1,1,4) 
        & \frac{569}{420} & \frac{149}{210} & -\frac{361}{1680}
        \\
        (1,1,1,5) 
        & \frac{629}{420} & \frac{4}{5} & -\frac{1}{1680}
        \\
        (1,1,1,6) 
        & \frac{689}{420} & \frac{109}{210} & -\frac{121}{1680}
        \\
        (1,1,2,1) 
        & \frac{473}{420} & \frac{53}{210} & -\frac{1129}{1680}
        \\
        (1,1,2,2) 
        & \frac{533}{420} & \frac{113}{210} & -\frac{1009}{1680}
        \\
        (1,1,2,3) 
        & \frac{593}{420} & \frac{173}{210} & -\frac{1369}{1680}
        \\
        (1,1,2,4) 
        & \frac{653}{420} & 1 & -\frac{529}{1680}
        \\
        (1,1,2,5) 
        & \frac{713}{420} & \frac{197}{210} & -\frac{169}{1680}
        \\
        (1,1,2,6) 
        & \frac{773}{420} & \frac{4}{7} & -\frac{289}{1680}
        \\
        (1,2,1,1) 
        & \frac{247}{210} & \frac{37}{105} & -\frac{109}{420}
        \\
        (1,2,1,2) 
        & \frac{277}{210} & \frac{67}{105} & -\frac{289}{420}
        \\
        (1,2,1,3) 
        & \frac{307}{210} & \frac{4}{5} & -\frac{169}{420}
        \\
        (1,2,2,1) 
        & \frac{289}{210} & \frac{4}{7} & -\frac{361}{420}
        \\
        (1,2,2,2) 
        & \frac{319}{210} & \frac{109}{105} &- \frac{121}{420}
        \\
        (1,2,2,3) 
        & \frac{349}{210} & \frac{139}{105} & -\frac{1}{420}
      \end{array}
    \end{equation*}
    \caption{
      $\mathcal{M}=\Sigma(3,4,5,7)$:
      We give
      the SU(2) irreducible representations of the fundamental group
      which contribute to the asymptotics of the WRT invariant
      $Z_N(\mathcal{M})$.
      None of the generators of the fundamental group~\eqref{fundamental_group}
      is  mapped to $\pm \id$.
    }
    \label{tab:3457}
  \end{table}

In Table~\ref{tab:3457_asymptotic}, we compute numerically both the
exact value~\eqref{Rozansky}
and the asymptotic value~\eqref{asymptotic_WRT_Seifert_4}
of the WRT invariant.
We can   see a good agreement also in this case.

\begin{table}[htbp]
  \centering
  \begin{equation*}
    \def\arraystretch{1.03}
    \begin{array}{r||cc}
      N & \text{exact result for $Z_N$} &
      \text{asymptotic formula
        $
        \begin{cases}
          (N+2) \cdot Z_N^{(0)} + Z_N^{(1)} 
          \\
          (N+2) \cdot Z_N^{(0)}
        \end{cases}
        $}
      \\
      \hline \hline
      998 &
      170.573359-
      7.19243844 \, \mathrm{i}
      &
      170.573296 -
      7.19236552 \, \mathrm{i}
      \\
      & &
      170.574879-
      7.19150956 \, \mathrm{i}
      \\
      999 &
      1.981255663-
      0.539792723 \, \mathrm{i}
      &
      1.981192840-
      0.539719917 \, \mathrm{i}
      \\
      & &
      2.018358388-
      0.430004395 \, \mathrm{i}
      \\
      1000 &
      10.9510287+
      54.4329988 \, \mathrm{i}
      &
      10.9509659+
      54.4330715 \, \mathrm{i}
      \\
      & &
      10.9917059+
      54.5856899 \, \mathrm{i}
      \\
      1001 &
      124.831004-
      123.107089 \, \mathrm{i}
      &
      124.830941 -
      123.107017 \, \mathrm{i}
      \\
      & &
      124.881006 -
      123.189839 \, \mathrm{i}
      \\
      1002 &
      61.4397540-
      97.5216937 \, \mathrm{i}
      &
      61.4396912-
      97.5216212 \, \mathrm{i}
      \\
      & &
      61.2653019 -
      97.4592987 \, \mathrm{i}
      \\
      1003 &
      146.275643+
      49.1637851 \, \mathrm{i}
      &
      146.275580+
      49.1638575 \, \mathrm{i}
      \\
      & &
      146.394939+
      49.4235715 \, \mathrm{i}
      \\
      1004 &
      172.965961+
      71.0176797 \, \mathrm{i}
      &
      172.965898+
      71.0177519 \, \mathrm{i}
      \\
      & &
      173.332924+
      71.1299446 \, \mathrm{i}
      \\
      1005 &
      99.17707586-
      1.446353766 \, \mathrm{i}
      &
      99.17701318-
      1.446281609 \, \mathrm{i}
      \\
      & &
      99.07532715-
      1.557181709 \, \mathrm{i}
      \\
      1006 &
      3.371067321-
      1.677974019 \, \mathrm{i}
      &
      3.371004661-
      1.677901969 \, \mathrm{i}
      \\
      & &
      3.385944423-
      1.713271357 \, \mathrm{i}
      \\
      1007 &
      67.04084613-
      35.56880740 \, \mathrm{i}
      &
      67.04078349-
      35.56873545 \, \mathrm{i}
      \\
      & &
      67.08289485-
      35.47827521 \, \mathrm{i}
    \end{array}
  \end{equation*}
  \caption{Numerical values of  the WRT invariant
    $Z_N(\mathcal{M})$ for
    $\mathcal{M}=\Sigma(3,4,5,7)$.
    We have used $P=420$ and $\phi=\frac{961}{420}$ in
    eq.~\eqref{Rozansky}.
  }
  \label{tab:3457_asymptotic}
\end{table}

\section{Discussions}

We have
studied the asymptotic expansion of the WRT
invariant for the Seifert manifold with 4-singular fibers.
A key is that the WRT invariant can be rewritten in terms
of a limiting value of the Eichler integrals of the modular forms with
half-integral weight.
A close connection between the quantum invariant and the modular form
was first observed in
Ref.~\citen{LawrZagi99a} for a case of the Poincar{\'e} homology
sphere.

In the case of  3-singular fibers,
we  showed in our previous paper~\cite{KHikami04b}
that the number of the non-vanishing Eichler integral coincides with
the number of the integral lattice points inside the 3-dimensional
tetrahedron.
As in known from Ref.~\citen{NeumWahl90a},
the number of the irreducible representations of the
fundamental group is related to that of the lattice points,
and 
this number is  proportional to the Casson invariant.
In this paper, as a generalization to
the case of  4-singular fibers,
we have shown that the
number of the lattice points in the 4-dimensional simplex
is related to the number of the non-vanishing Eichler integrals.
We have also clarified  a relationship with the representation of the 
fundamental group.

Our modular form with half-integral weight 
can be easily generalized as follows.
We fix $M$-tuple
$\vv{p}=(p_1, p_2, \dots, p_M)$, where $p_i\geq 2$ are pairwise
coprime positive integers.
We set $P= \prod_{j=1}^M p_j$, and  define the periodic function
by
\begin{equation}
  \chi_{2 P}^{\vv{\ell}}(n)
  =
  \begin{cases}
    \displaystyle
    - \prod_{j=1}^M \varepsilon_j
    &
    \text{if
      $\displaystyle
      n \equiv
      P \, \left(
        1 + \sum_{j=1}^M \varepsilon_j \,
        \frac{\ell_j}{p_j}
      \right)
      \mod 2 \, P
      $
    }
    \\[2mm]
    0 &
    \text{others}
  \end{cases}
\end{equation}
where
$\varepsilon_j\in\{1, -1\}$, and
$\vv{\ell}=(\ell_1, \dots, \ell_M) \in \mathbb{Z}^M$
is an $M$-tuple
satisfying
$0<\ell_j< p_j$.
The case of $M=2$  appears as the Virasoro character of the
minimal model.
There are $2^M$ non-zero terms
$\chi_{2 P}^{\vv{\ell}}(n)$
in one period
$n\in[0, 2 \, P]$, and we have a mean value zero
\begin{equation*}
  \sum_{n=0}^{2 \, P-1} \chi_{2 P}^{\vv{\ell}}(n) = 0
\end{equation*}
One sees that the function $\chi_{2 P}^{\vv{\ell}}(n)$ is even
(resp. odd)
if $M$ is even (resp. odd).
When we define the functions $\Phi_{\vv{p}}^{\vv{\ell}}(\tau)$ by
\begin{equation}
  \Phi_{\vv{p}}^{\vv{\ell}}(\tau)
  =
  \begin{cases}
    \displaystyle
    \frac{1}{2 } \, \sum_{n \in \mathbb{Z}} \chi_{2 P}^{\vv{\ell}}(n) \,
    q^{\frac{n^2}{4 \, P}}
    &
    \text{if $M$ is even}
    \\[2mm]
    \displaystyle
    \frac{1}{2 } \, \sum_{n \in \mathbb{Z}}
    n \, \chi_{2 P}^{\vv{\ell}}(n) \,
    q^{\frac{n^2}{4 \, P}}
    &
    \text{if $M$ is  odd}
  \end{cases}
\end{equation}
we find that it 
is a vector modular form with dimension
\begin{equation*}
  D= \frac{1}{2^{M-1}} \, \prod_{j=1}^M (p_j - 1 )
\end{equation*}
due to the symmetry under the involutions $\sigma_{i,j}$,
and that the weight is $1/2$
(resp. $3/2$) when $M$ is even (resp. odd).
Generally the SU(2)
WRT invariant for the Seifert manifold
$\Sigma(p_1,\dots,p_M)$
with $M$-singular
fibers is  related to this vector modular form.

As a generalization of the correspondence between the lattice points
and the non-vanishing Eichler integrals,
we propose the following
conjecture.
\begin{conj}
  If
  we set
  $\gamma(\vv{p})$   as the number of $M$-tuples $\vv{\ell}$ satisfying
  \begin{equation*}
    \sum_{n=1}^{2 \, P}
    \chi_{2 P}^{\vv{\ell}}(n) \,
    B_{M-2} \left( \frac{n}{2 \, P} \right)
    \neq 0
  \end{equation*}
  then $D-\gamma(\vv{p})$ coincides with the
  number of the lattice points satisfying
  \begin{equation*}
    0<\sum_{j=1}^M \frac{\ell_j}{p_j} < 1
  \end{equation*}
\end{conj}
A case of $M=2$ is trivial (see Ref.~\citen{KHikami03c}),
as we have $\gamma=0$ due to $B_0(x)=1$ and the  definition of
$\sum_{n=1}^{2 P} \chi_{2 P}^{\vv{\ell}}(n)=0$.
A case of $M=3$ was shown in  Ref.~\citen{KHikami04b},
and we have proved this conjecture for $M=4$ in this paper.

\section*{Acknowledgments}
The author would like to thank J.~Kaneko, A. N. Kirillov,
and H.~Murakami for useful communications.
This work is supported in part  by Grant-in-Aid for Young Scientists
from the Ministry of Education, Culture, Sports, Science and
Technology of Japan.


\end{document}